\documentclass[12pt]{iopart}
\usepackage{iopams}
\usepackage{graphicx}

\begin{document}
\title[Constraining a matter-dominated cosmological model with
constant bulk viscous] {Can a matter-dominated model with constant
bulk viscosity drive the accelerated expansion of the universe?}

\author{Arturo Avelino and Ulises Nucamendi}

\address{Instituto de F\'{\i}sica y Matem\'aticas\\
Universidad Michoacana de San Nicol\'as de Hidalgo \\
Edificio C-3, Ciudad Universitaria, CP. 58040\\
Morelia, Michoac\'an, M\'exico\\ }

\eads{\mailto{avelino@ifm.umich.mx}, \mailto{ulises@ifm.umich.mx}}

\begin{abstract}
We test a cosmological model which the only component is a
pressureless fluid with a constant bulk viscosity as an explanation
for the present accelerated expansion of the universe. We classify
all the possible scenarios for the universe predicted by the model
according to their past, present and future evolution and we test
its viability performing a Bayesian statistical analysis using the
SCP ``Union'' data set (307 SNe Ia), imposing the second law of
thermodynamics on the dimensionless constant bulk viscous
coefficient $\tilde{\zeta}$ and comparing the predicted age of the
universe by the model with the constraints coming from the oldest
globular clusters.

The best estimated values found for $\tilde{\zeta}$ and the Hubble
constant $H_0$ are: $\tilde{\zeta}=1.922 \pm 0.089$ and $H_0=69.62
\pm 0.59 \;({{\rm km}}/{{\rm s}}){{\rm Mpc}}^{-1}$ with a
$\chi^2_{\rm min}=314 \;$ ($\chi^2_{\rm{d.o.f}}=1.031$). The age of
the universe is found to be $14.95  \pm 0.42$ Gyr. We see that the
estimated value of  $H_0$ as well as of $\chi^2_{\rm{d.o.f}}$ are
very similar to those obtained from $\Lambda$CDM model using the
same SNe Ia data set. The estimated age of the universe is in
agreement with the constraints coming from the oldest globular
clusters. Moreover, the estimated value of $\tilde{\zeta}$ is
positive in agreement with the second law of thermodynamics (SLT).

On the other hand, we perform different forms of marginalization
over the parameter $H_0$ in order to study the sensibility of the
results to the way how $H_0$ is marginalized. We found that it is
almost negligible the dependence between the best estimated values
of the free parameters of this model and the way how $H_0$ is
marginalized in the present work.

Therefore, this simple model might be a viable candidate to explain
the present acceleration in the expansion of the universe.

\end{abstract}
 \submitto{Journal of Cosmology
and Astroparticle physics} \maketitle

        \section{Introduction.}

In the last ten years the observations of type Ia supernovae (SNe Ia) have suggested a
possible late time accelerated expansion of the universe (see for instance
\cite{Riess1998}-\cite{Wood-Vasey2007} and references therein).
This discovery has additional support from the cosmic microwave background \cite{Bennett2003}
 and the large scale structure \cite{Tegmark2004} observations.

Several models have been proposed to explain this recent
acceleration, one of them is the assumption of the existence of the
so-called \emph{dark energy} as responsible of such acceleration,
i.e.,  a new unknown component that must be $\sim 70 \% $ of the
total content of matter and energy in the universe
\cite{Riess1998}-\cite{Wood-Vasey2007}. The leading dark energy
candidates are a cosmological constant and a slowly varying rolling
scalar field (quintessence models)
\cite{Caldwell1998}-\cite{Peebles2003}. However, a cosmological
constant model faces several strong problems, one of them is the
huge discrepancy between its predicted and observed value  (of about
120 orders of magnitude) \cite{Weinberg1989}-\cite{Padmanabhan},
another one is the so-called the ``\emph{cosmic coincidence
problem}'', i.e., why are we living in a moment where the matter
density in the universe is of the same order than the dark energy
density? \cite{Steinhardt1997}-\cite{Steinhardt1999}.

On the other hand, it has been known since several years ago before
of the discovery of the present acceleration in the expansion of the
universe that a bulk viscous fluid can produce an accelerating
cosmology (although it was originally proposed in the context of an
inflationary period in the early universe) without the need of any
cosmological constant or dark energy component
\cite{Heller1973}-\cite{Maartens1997} although some authors do not
agree with this conclusion \cite{Hiscock1991}.

So, it is natural to think of the bulk viscous pressure as one of
the possible mechanism that can accelerate the  universe today (see
for instance \cite{Kremer2003}-\cite{Meng2008}, where in some of
these works the bulk viscous coefficient is assumed a priori without
being derived from known physics or known particle properties).
However, this idea faces the problem of that it is necessary to
propose a viable mechanism for the origin of the bulk viscosity and
in this sense some proposals have been already suggested
\cite{Zimdahl2000}-\cite{Mathews2008}.

Nowadays, with the observational data sets of SNe Ia
\cite{Riess1998}-\cite{Wood-Vasey2007} and some other cosmological
observations like the shift parameter $R$ of the cosmic microwave
background radiation (CMB) given by the Wilkinson Microwave
Anisotropy Probe (WMAP) observations \cite{Mukherjee2006}, and the
baryon acoustic oscillation (BAO) measurement from the Sloan Digital
Sky Survey (SDSS) \cite{Eisenstein} it is possible to \textit{test}
cosmological models. So, in the present work we study and test bulk
viscous matter-dominated cosmologies, i.e., scenarios which the only
component of the universe is a pressureless fluid with bulk
viscosity. The pressureless fluid characterizes to both baryon and
dark matter components. The idea of this model is to explain the
present acceleration of the universe using the bulk viscous pressure
of the fluid without the need of any dark energy.

As it was mentioned above, the explicit form of the bulk viscosity
has to be assumed a priori or obtained from a known physical
mechanism. In the present work we choose the first possibility. So,
we assume a \emph{constant} bulk viscous coefficient that it is the
simplest parametrization for the bulk viscosity and we estimate its
value using SNe Ia observations. The theoretical solution of this
model has been also analyzed in \cite{Heller1973},
\cite{Klimek1974}-\cite{ArturoUlisesProc1} and other
parametrizations for the bulk viscous coefficient $\zeta$ have been
also proposed and studied in \cite{Kremer2003}-\cite{Singh2007} and
\cite{Barrow1987}-\cite{ArturoUlisesProc2}. There is also other
 general approach to the bulk viscous cosmologies called
``fluids with inhomogeneous equation of state'' (see
\cite{Nojiri2005,Stefancic2008B}).

In \sref{SectionBulkViscosityTheory} we present the generalities of
bulk viscous  fluids in General Relativity (GR), in
\sref{SectionBulkViscousMatterDominatedModel} we apply this
formalism to a bulk viscous matter-dominated universe model  where
we find the explicit expression for the Hubble parameter in function
of the redshift. In \sref{SectionScaleFactor} we analyze the
behavior of the scale factor  for the possible scenarios that the
model predicts for the universe according to the value of the bulk
viscous coefficient. In \sref{SectionCosmologicalParameters} we
study the behavior of the deceleration parameter $q$, the curvature
scalar $R$, the total matter density $\rho_{\rm m}$ and the age of
the universe and in \sref{Sectionthermodynamics} we briefly review
the second law of thermodynamics (SLT). \Sref{SectionSNeTest}
presents the SNe Ia test to constrain the model and compute the best
estimated values for the bulk viscous coefficient and the Hubble
constant. After that, \sref{SectionMarginalizationHo} presents the
best estimated values and  the probability distribution functions
for the bulk viscous coefficient by marginalizing over the Hubble
constant assuming three different priors and finally in
\sref{SectionConclusions} we present our conclusions. The appendices
A, B and C detail the marginalization methods used in the present
work.

        \section{Theory of relativistic bulk viscous fluids.}
        \label{SectionBulkViscosityTheory}

The origin of the bulk viscosity in a physical system is due to its
deviations from the local thermodynamic equilibrium (for a review in
the theory of relativistic dissipative fluids see
\cite{MaartensHanno}). In a cosmological fluid, the bulk viscosity
arises when the fluid expands (or contracts) too fast so that the
system does not have enough time to restore the local thermodynamic
equilibrium and then it appears an \emph{effective pressure}
restoring the system to its local thermodynamic equilibrium. The
bulk viscosity can be seen as a measurement of this effective
pressure. When the fluid reaches again the thermal equilibrium then
the bulk viscosity pressure ceases \cite{Wilson},
\cite{Okumura}--\cite{Xinzhong}.

In an \textit{accelerated} expanding universe it is very possible
that the expansion process is actually a collection of states out of
thermal equilibrium in a small fraction of time. So, it is natural
to assume the existence of a bulk viscous coefficient in a more
realistic description of the accelerated universe today.

We use the Weinberg formalism \cite{Weinberg}--\cite{Hofmann} for
the imperfect fluids. So, in the present work we consider a bulk
viscous fluid as source of matter in the Einstein fields equations
$G_{\mu \nu} = 8\pi GT_{\mu \nu}$, where $G$ is the Newton
gravitational constant.

On the other hand, the energy-momentum tensor of an imperfect fluid
with a first-order deviation from the thermodynamic equilibrium has
the form \cite{Wilson,Weinberg,Misner}:
\begin{equation}\label{Energy_momentum_tensor}
    T_{\mu\nu}=\rho u_\mu u_\nu + (g_{\mu\nu}+u_\mu u_\nu)P^*
\end{equation}
\noindent where
\begin{equation}\label{pressure}
    P^* \equiv P - \zeta \nabla^{\nu}u_{\nu}
\end{equation}

\noindent In the equations \eref{Energy_momentum_tensor} and
\eref{pressure} the four-velocity vector $u_\nu$ is of an observer
who measures the \textit{effective} pressure $P^*$, $P$ and $\rho$
are the pressure and density of the fluid respectively. The term
$\zeta$ is the bulk viscous coefficient that arises in the fluid
which is out of the local thermodynamic equilibrium.

It can be seen that the energy-momentum tensor
\eref{Energy_momentum_tensor} is similar to that of a perfect fluid
but with an effective pressure $P^*$ composed by the usual pressure
$P$ of the fluid plus the pressure due to its bulk viscosity
$P_{{\rm visc}} \equiv -\zeta \nabla^{\nu}u_{\nu}$. This viscous
pressure, $P_{{\rm visc}}$, can be seen as a ``measurement'' of the
pressure to restore the local thermodynamic equilibrium
\cite{Wilson}, \cite{Okumura}--\cite{Xinzhong}. The conservation
equation for the viscous fluid is
\begin{equation}\label{equation1PerfectFluid}
    u^{\nu} \nabla_{\nu} \rho + (\rho + P^*) \nabla^{\nu} u_{\nu} =0
\end{equation}

The effective pressure \eref{pressure} was originally proposed by
Eckart \cite{Eckart1940} in 1940 for a relativistic dissipative
process in the context of thermodynamics systems out of local
equilibrium, and subsequently  Landau \& Lifshitz presented an
equivalent formulation \cite{Landau}.

However, the Eckart theory has problems in several aspects. One of
them is that all the equilibrium states in this theory are unstable
\cite{Hiscock1985}. Another one is that signals can propagate
through the fluids with superluminal velocities
\cite{Muller1967,Israel1976}.

In 1979, Israel-Stewart \cite{IsraelStewart1979A,IsraelStewart1979B}
developed a more consistent and general theory that avoids these
problems, and from which the Eckart theory is the first-order limit
of the Israel-Stewart theory when the relaxation time goes to zero.
Nevertheless, the Eckart theory is simpler to deal with than the
Israel-Stewart theory.

Despite of the inherent problems of the Eckart theory, and due to
that it is simpler than the Israel-Stewart theory, it has been
widely used recently by several authors to model  bulk viscous dark
fluids as responsible of the recent observed acceleration of the
universe assuming that the approximation is valid for this purpose
(see, for instance \cite{Kremer2003}-\cite{Jie2006-B},
\cite{Colistete2007,Singh2007,Wilson,ArturoUlisesProc1,ArturoUlisesProc2}),
i.e., in these papers assume a vanishing relaxation time, so that,
in this limit the Eckart theory is a good approximation to the
Israel-Stewart theory. In this context, it is convenient to mention
that Hiscock \etal \cite{Hiscock1991} showed that flat
Friedmann-Robertson-Walker cosmological models containing a bulk
viscous Boltzmann gas expand more rapidly using the Eckart theory
than the Israel-Stewart one, and suggesting that inflationary
acceleration driven by bulk viscosity could be an effect of applying
a pathological theory such as the Eckart theory. These results
suggest that in this context the use of the Israel-Stewart causal
theory would not produce a recent accelerating epoch as the Eckart
theory would. However, posterior studies have suggested that this
conclusion could not be true because consistent inflationary
solutions have been found using the Israel-Stewart theory
\cite{Zimdahl1996,Maartens1997}.

It is important to point out that there exists other more general
formulation for irreversible processes than the Israel-Stewart
theory developed by D. Pav\'on et al. where the temperature is not
necessary that of the thermodynamic equilibrium, see
\cite{Pavon1982} for details. In the present work we assume the
Eckart theory and we constrain the model described above using SNe
Ia observations.

\section{Cosmological model of  bulk viscous matter-dominated universes.}
        \label{SectionBulkViscousMatterDominatedModel}

We study a cosmological model in a flat universe where the only
component is a pressureless fluid with constant bulk viscosity  as
an explanation for the present accelerated expansion of the
universe. The pressureless fluid characterizes both the baryon and
dark matter components.

Note that this model does not have the Cosmic Coincidence problem
and that this fluid represents an unified description of the dark
sector plus the baryon component in a similar way than the Chaplygin
gas model (see for instance \cite{Colistete2007,Jamil2008B} and
references therein). In this approach the present acceleration of
the universe is driven by the bulk viscous pressure of the fluid
instead of a dark energy component. The present work differs from
the approach of Colistete \etal \cite{Colistete2007} in that they
propose two fluids, one of them is a bulk viscous fluid representing
in an unified way the dark sector with a bulk viscous coefficient
$\zeta$ proportional to a power of the energy density $\rho$ (i.e.,
$\zeta=\tilde{\zeta} \; \rho^{\nu}$, with $\tilde{\zeta}, \nu$
constants), and the other one is a pressureless fluid representing
the baryon component. Nevertheless, most of their analysis and test
using cosmological observations is done using the ansatz $\nu=0$.

Since we work with a pressureless fluid ($P=0$) then $P^* =P_{{\rm
visc}} \equiv - \zeta \nabla^{\nu}u_{\nu}$, where $\zeta$ is the
bulk viscous coefficient of the matter fluid.

We consider a spatially flat geometry for the
Friedmann-Robertson-Walker (FRW) cosmology as favored by WMAP
\cite{WMAP}
\begin{equation}\label{FRM_metric}
    ds^2=-dt^2+a^2(t)(dr^2+r^2d\Omega^2)
\end{equation}
where the function $a(t)$ is the \emph{scale factor}. On the other
hand, we spread the conservation equation
\eref{equation1PerfectFluid} in all its components
\begin{equation}\label{ConservationEquation}
\dot{\rho}_{{\rm m}} + (\rho_{{\rm m}} - 3H\zeta)
 3H  =0
\end{equation}

\noindent where $H \equiv \dot{a}/a$ is the \emph{Hubble parameter},
$\rho_{\rm m}$ is the total matter density, the dot means time
derivative and $\nabla^{\nu} u_{\nu}=3H$.

The conservation equation \eref{ConservationEquation} can be written
in terms of the scale factor as
\begin{equation}\label{conservation_equationMatter0_ODE}
    a\frac{d\rho_{{\rm m}}}{da} = 3\left( 3H \zeta - \rho_{{\rm m}} \right)
\end{equation}
This equation is valid for any parametrization of $\zeta$, in
particular we assume the ansatz  $\zeta=$ constant that it is
perhaps the simplest parametrization for the bulk viscous
coefficient that can be proposed. Its value is going to be estimated
from the SNe Ia observations.

On the other hand, the first Friedmann equation for a flat universe is
\begin{eqnarray}
\label{FriedmannEq1Matter}
H^2 = \frac{8\pi G}{3}\rho_{{\rm m}}
\end{eqnarray}
So, substituting \eref{FriedmannEq1Matter} into
\eref{conservation_equationMatter0_ODE} we obtain
\begin{equation}\label{conservation_equation2Matter0_ODE}
    a\frac{d\rho_{{\rm m}}}{da} + 3\rho_{{\rm m}} - \gamma \rho_{{\rm m}}^{1/2}=0,
    \quad {\mbox {where}} \quad \gamma \equiv
    9 \left(\frac{8 \pi G}{3}\right)^{1/2} \zeta
\end{equation}

Changing of variable from the scale factor to the redshift $z$ using
the relationship\footnote{We assume that the value of the scale
factor evaluated \emph{today} is equal to one.} $a = 1/(1+z)$, we
obtain the ordinary differential equation (ODE):

\begin{equation}\label{conservation_equationZMatter0_ODE}
    (1+z)\frac{d\rho_{{\rm m}}}{dz} - 3\rho_{{\rm m}} + \gamma
    \rho_{{\rm m}}^{1/2}=0
\end{equation}

The exact solution of this ODE is:
\begin{equation}\label{SolutionToODE-Z0}
    \rho_{{\rm m}}(z)= \left[ \frac{\gamma}{3}+ \left(
    \rho^{1/2}_{{\rm m}0} -\frac{\gamma}{3} \right) (1+z)^{3/2} \right]^2
\end{equation}
where $\rho_{{\rm m}0}$ is the matter density evaluated today.
Substituting this solution into \eref{FriedmannEq1Matter} we obtain

\begin{equation}\label{SolutionToODE-b-Z0}
H^2(z)= H_0^2 \left[\frac{\tilde{\zeta}}{3} +
 \left( \Omega^{1/2}_{{\rm m}0} -
\frac{\tilde{\zeta}}{3} \right)(1+z)^{3/2} \right]^2
\end{equation}
where $H_0$ is the Hubble constant and where we have defined the dimensionless bulk
viscous coefficient $\tilde{\zeta}$, the matter density parameter $\Omega_{{\rm m}0}$ and the
critical density today $\rho^0_{{\rm crit}}$ as:
\begin{equation}\label{DefinitionsZoBar}
\tilde{\zeta} \equiv \frac{24\pi G}{H_0} \zeta, \qquad
\Omega_{{\rm m}0} \equiv \frac{\rho_{{\rm m}0}}{\rho^0_{{\rm crit}}}, \qquad
\rho^0_{{\rm crit}} \equiv \frac{3H_0^2}{8 \pi G}
\end{equation}

In this model the bulk viscous matter is the only  component of the
universe implying  that the first Friedmann equation
\eref{FriedmannEq1Matter} evaluated today is $\Omega_{{\rm m}0}=1$.
With this, the expression \eref{SolutionToODE-b-Z0} finally becomes
\begin{equation}\label{SolutionFinalToODE-Z0}
H(z)= \frac{H_0}{3} \left[\tilde{\zeta} + \left(3
-\tilde{\zeta} \right)(1+z)^{3/2} \right]
\end{equation}

\section{Classification and evolution of bulk viscous matter-dominated models.}
            \label{SectionScaleFactor}

In this section we classify all the possibilities for the universe
predicted by this bulk viscous matter-dominated model using
different values of the constant bulk viscous coefficient. Note that
we have one different model for each value of $\tilde{\zeta}$, so
actually we have a collection of models depending of the value of
$\tilde{\zeta}$.

We analyze  the theoretical behavior of the scale factor in terms of
the cosmic time. We begin expressing \eref{SolutionFinalToODE-Z0} in
terms of the scale factor

\begin{equation}\label{HubbleParameterAZ0}
  H(a)\equiv \frac{\dot{a}}{a}=  \frac{H_0}{3}
\left(  \frac{\tilde{\zeta} a^{3/2}+3-\tilde{\zeta}}{a^{3/2}} \right)
\end{equation}

\noindent Integrating \eref{HubbleParameterAZ0} yields

\begin{equation}
H_0(t-t_0)=H_0\int^t_{t_0} dt' = 3 \int^a_{1}
\frac{a'^{1/2} da'}{\tilde{\zeta} a'^{3/2}+3-\tilde{\zeta}}
\label{ScaleFactorIntegrals}
\end{equation}
\noindent where $t$ labels the \emph{cosmic time} and $t_0$ the
cosmic time \emph{today}.

            \subsection{Case $\tilde{\zeta}=0$.}
            \label{SectionZ0eq0}

When $\tilde{\zeta}=0$ we recover the usual matter-dominated
universe (with null bulk viscosity) with a scale factor coming from
the integration of \eref{ScaleFactorIntegrals} like:

\begin{equation}
a(t)= \left(  \frac{3}{2}H_0(t-t_0)+1\right)^{2/3}, \quad \qquad \tilde{\zeta} = 0
\label{ScaleFactorForZ0eq0Z0}
\end{equation}
The first and second derivatives of the scale factor with respect to $x\equiv H_0(t-t_0)$
are

\begin{equation}\label{ScaleFactorDeriveZ0eq0}
    \frac{da}{dx}= \left( \frac{2}{3x+2 }\right)^{1/3}
\end{equation}
and
\begin{equation}\label{ScaleFactor2ndDeriveZ0eq0}
    \frac{d^2a}{dx^2} = - \frac{2^{1/3}}{\left(3x +2\right)^{4/3}}
\end{equation}

The plot of functions
\eref{ScaleFactorForZ0eq0Z0}--\eref{ScaleFactor2ndDeriveZ0eq0} are
shown in figures
\ref{ScaleFactorFor0z3Plot}--\ref{ScaleFactor2ndDerivedadxPlot}
respectively (the long dashed lines).

This case  predicts an \emph{eternal decelerated  expanding}
universe (see figures
\ref{ScaleFactorFor0z3Plot}--\ref{DecelerationParameterPlot}). In
this case the curvature scalar  and the matter density  are
$R=(3H_0^2)a^{-3}$ and $\rho_{\rm m}= (3H_0^2/ 8\pi G) a^{-3}$
respectively, that both diverge when the scale factor goes to zero
(see sections \ref{SectionScalarOfCurvatureZ0} and
\ref{SectionDensityMatterZ0} for details). When we have these two
conditions ($R(a\rightarrow0)\rightarrow \infty$ and $\rho_{\rm
m}(a\rightarrow0) \rightarrow \infty$) in the past of the universe
we say that there was a \emph{Big-Bang}. The elapsed time between
the Big-Bang time till today is
\begin{equation}\label{BigBangTimeZ0eq0}
t_{\rm B}= t_0 - \frac{2}{3 H_0}
\end{equation}
where the subscript `B' stands for ``Big-Bang''. On the other hand,
when $a \rightarrow \infty$ then $R$ and $\rho_{\rm m}$ decrease to
zero.

            \subsection{Case $\tilde{\zeta}\neq0$.}
            \label{SectionZ0neq0}

For $\tilde{\zeta} \neq0$ we do the change of variable $y \equiv
a^{3/2}$ in expression \eref{ScaleFactorIntegrals} yielding:
\begin{equation}
H_0 (t-t_0)= \frac{2}{\tilde{\zeta}} \int^y_{1}
\frac{\tilde{\zeta} dy'} {\tilde{\zeta} y'+3-\tilde{\zeta}} =
\frac{2}{3 \tilde{\zeta}} \ln \left|\tilde{\zeta} a^{3/2}
+3-\tilde{\zeta} \right|
\end{equation}
This last expression can be rewritten as
\begin{equation}
3 \exp \left[ \frac{\tilde{\zeta}}{2} H_0(t-t_0) \right] = |\tilde{\zeta} a^{3/2} +
3-\tilde{\zeta}|
\label{ScaleFactorEquation}
\end{equation}
When $\tilde{\zeta} a^{3/2} + 3-\tilde{\zeta} \geq0$ we can remove the absolute value bar
of the right hand side term and then we obtain

\begin{equation}\label{ScaleFactorFor0z3}
    a(t)= \left[ \frac{3 \exp \left( \frac{1}{2} \tilde{\zeta}
H_0(t-t_0) \right) -3 + \tilde{\zeta}}{\tilde{\zeta}}
\right]^{2/3}, \qquad \tilde{\zeta} \neq 0
\end{equation}
The first and second derivative of the scale factor with respect to $x \equiv H_0(t-t_0)$
are

\begin{equation}\label{ScaleFactorDeriveZ0}
    \frac{da}{dx}= \left[ \frac{\tilde{\zeta} \; \exp
    \left( \frac{3}{2} \tilde{\zeta} x \right)}{3\exp
    \left( \frac{3}{2} \tilde{\zeta} x \right)+\tilde{\zeta}-3} \right]^{1/3}
\end{equation}
and
\begin{equation}\label{ScaleFactor2ndDeriveZ0}
    \frac{d^2a}{dx^2}= \frac{\tilde{\zeta}^{4/3} \; \exp
    \left( \frac{1}{2} \tilde{\zeta} x \right) \left[ 2\exp
    \left( \frac{1}{2} \tilde{\zeta} x \right)+\tilde{\zeta}-3 \right] }{2\left[3\exp
    \left( \frac{1}{2} \tilde{\zeta} x \right)+\tilde{\zeta}-3\right]^{4/3}}
\end{equation}
The behavior of the expressions \eref{ScaleFactorFor0z3}--\eref{ScaleFactor2ndDeriveZ0}
are shown in figures \ref{ScaleFactorFor0z3Plot}--\ref{ScaleFactor2ndDerivedadxPlot}
respectively.

The case $\tilde{\zeta} a^{3/2} + 3-\tilde{\zeta} < 0$ does not correspond to a physical
case of our interest because this case predicts an eternal \emph{contraction} of the
universe ($H(t)<0$ for any value of the cosmic time $t$) that is in contradiction with
the observations (see the expression \eref{HubbleParameterAZ0} to note that
$\tilde{\zeta} a^{3/2} + 3-\tilde{\zeta} < 0$ implies $H(t)<0$).

In order to study carefully the expression \eref{ScaleFactorFor0z3} for different values
of $\tilde{\zeta}$ we take four cases:
\begin{itemize}
  \item $0< \tilde{\zeta} < 3$
  \item $\tilde{\zeta} = 3$
  \item $\tilde{\zeta}>3$
  \item $\tilde{\zeta} <0$
\end{itemize}

            \subsubsection{Case $0<\tilde{\zeta} < 3$.}
            \label{Section0Z03}

From the expression \eref{ScaleFactorFor0z3} we can see  that when
$t \rightarrow \infty$ then the scale factor tends to have the form
like that of the {\it de Sitter universe}, i.e.,

\begin{equation}\label{ScaleFactorProportionaldeSitter}
a(t) \propto {\rm e}^{(\tilde{\zeta}/3) H_0(t-t_0)}
\end{equation}

On the other hand, for any value of $\tilde{\zeta}$ in this interval
all the models predict an universe having a Big-Bang in the past in
the cosmic time:

\begin{equation}\label{BigBangTime0Zo3}
t_{\rm B}= t_0 + \frac{2}{\tilde{\zeta} H_0} \ln\left( 1- \frac{\tilde{\zeta}}{3} \right)
\end{equation}

\begin{figure}
\begin{center}
\includegraphics[width=12cm]{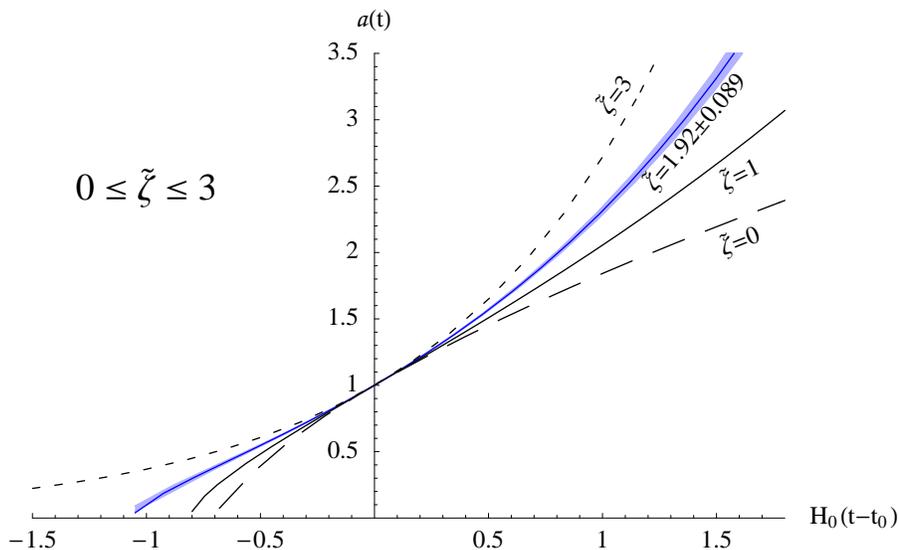}\\
\caption{Plot of $a(t,\tilde{\zeta})$ for different values of
$\tilde{\zeta}$ in the interval $0\leq\tilde{\zeta} \leq 3$ (see
expressions \eref{ScaleFactorForZ0eq0Z0} and
\eref{ScaleFactorFor0z3}). $H_0$ is the Hubble constant, $t$ is the
cosmic time and $t_0$ is its present value. The long dashed line
corresponds to $\tilde{\zeta}=0$ (a flat matter-dominated universe
with null bulk viscosity, see expression
\eref{ScaleFactorForZ0eq0Z0}). The short dashed line corresponds to
$\tilde{\zeta}=3$ (the \emph{de Sitter} universe). The blue line
with the blue band correspond to a model with $\tilde{\zeta}=
1.922\pm 0.089$. This is the best estimated value of $\tilde{\zeta}$
coming from the SCP ``Union'' SNe Ia data set analysis (see section
\ref{SectionSNeTest}). The band corresponds to the error at the
68.3\% confidence level.} \label{ScaleFactorFor0z3Plot}
\end{center}
\end{figure}

\begin{figure}
\begin{center}
\includegraphics[width=12cm]{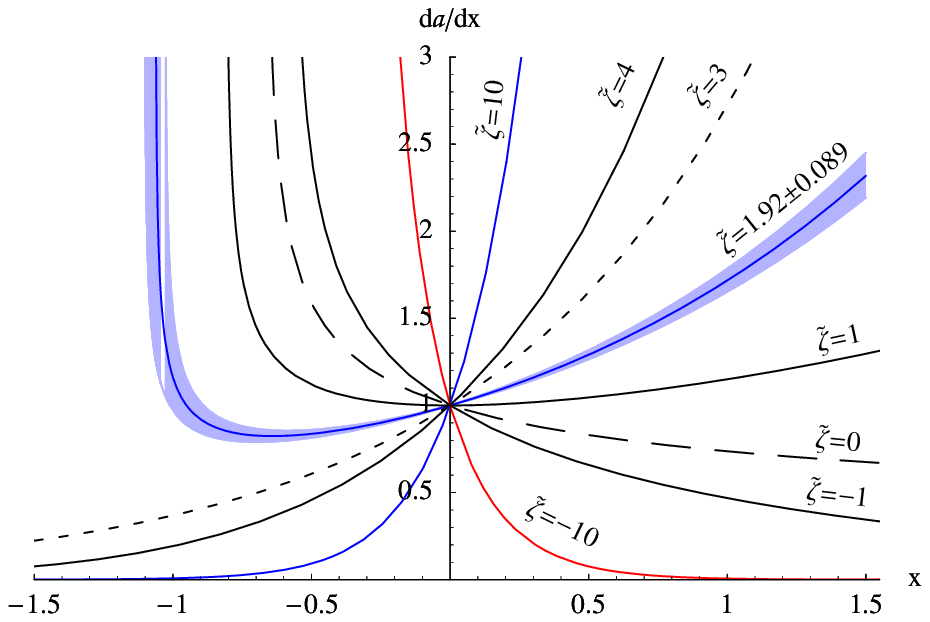}\\
\caption{Plot of the derivative of the scale factor with respect to
$x\equiv H_0(t-t_0)$ for different values of $\tilde{\zeta}$ (see
expression \eref{ScaleFactorDeriveZ0}). The long dashed line
corresponds to $\tilde{\zeta}=0$ (a flat matter-dominated universe
with null bulk viscosity, see expression
\eref{ScaleFactorDeriveZ0eq0}) and the short dashed line corresponds
to $\tilde{\zeta}=3$ (the \emph{de Sitter} universe). The blue line
with the blue band correspond to a model with $\tilde{\zeta}=
1.922\pm 0.089$. This is the best estimated value of $\tilde{\zeta}$
coming from the SCP ``Union'' SNe Ia data set analysis (see section
\ref{SectionSNeTest}). The band corresponds to the error at the
68.3\% confidence level.} \label{ScaleFactorDerivedadxPlot}
\end{center}
\end{figure}

\begin{figure}
\begin{center}
\includegraphics[width=8cm]{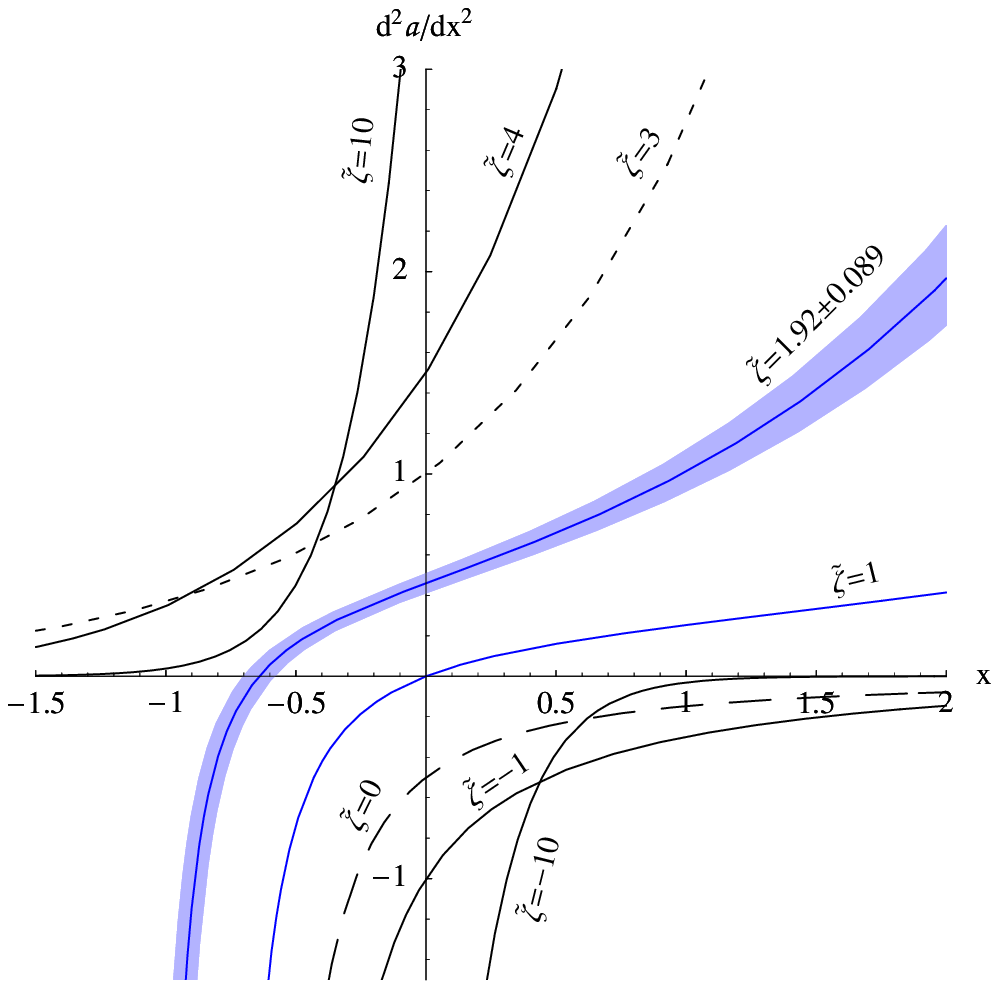}\\
\caption{Plot of the \emph{second} derivative of the scale factor
with respect to $x\equiv H_0(t-t_0)$ for different values of
$\tilde{\zeta}$ (see expression \eref{ScaleFactor2ndDeriveZ0}). The
long dashed line corresponds to $\tilde{\zeta}=0$ (a flat
matter-dominated universe with null bulk viscosity, see expression
\eref{ScaleFactor2ndDeriveZ0eq0}) and the short dashed line
corresponds to $\tilde{\zeta}=3$ (the \emph{de Sitter} universe).
The blue line with the blue band correspond to a model with
$\tilde{\zeta}= 1.922\pm 0.089$. This is the best estimated value of
$\tilde{\zeta}$ coming from the SCP ``Union'' SNe Ia data set
analysis (see section \ref{SectionSNeTest}). The band corresponds to
the error at the 68.3\% confidence level. We see that for
$\tilde{\zeta}\leq 0$ all the models predict a decelerated universe
($d^2a/dx^2<0$) \emph{forever}. For $\tilde{\zeta}\geq 3$ the models
predict an accelerated universe ($d^2a/dx^2>0$) \emph{forever}. And
in the range $0<\tilde{\zeta} < 3$ the models predict a transition
from deceleration to acceleration.}
  \label{ScaleFactor2ndDerivedadxPlot}
\end{center}
\end{figure}

So, the universe begins with a Big-Bang followed by an \emph{eternal
expansion} (there is not any recollapse epoch, see figures
\ref{ScaleFactorFor0z3Plot} and \ref{ScaleFactorDerivedadxPlot}) and
this expansion begins with a decelerated epoch followed by an
\emph{eternal accelerated} one (see figures
\ref{ScaleFactorDerivedadxPlot} and
\ref{ScaleFactor2ndDerivedadxPlot}).

The \emph{transition} between the decelerated--accelerated expansion
epochs depends on the value of $\tilde{\zeta}$. We compute the value
of the scale factor where the transition happens. For that, we
derive to $\dot{a}$ with respect to $a$ using the expression
\eref{HubbleParameterAZ0}:
\begin{equation}\label{ScaleFactorDeriveRespect-a-Z0}
    \frac{d \dot{a}}{d a}= \frac{H_0}{3}
    \left( \tilde{\zeta} + \frac{\tilde{\zeta}-3}{2a^{3/2}} \right)
\end{equation}
Then, we make equal to zero to expression
\eref{ScaleFactorDeriveRespect-a-Z0} in order to obtain the value of
the scale factor ``$a_{\rm t}$'' where the transition happens,
obtaining
\begin{equation}\label{ScaleFactorTransition0leqZleq3Z0}
 a_{\rm t} = \left( \frac{3-\tilde{\zeta}}{2\tilde{\zeta}} \right)^{2/3}
\end{equation}
where the subscript ``t'' stands for ``transition''. Writing the expression
\eref{ScaleFactorTransition0leqZleq3Z0} in terms of the redshift $z$ yields
\begin{equation}\label{RedshiftTransition0leqZleq3Z0}
 z_{\rm t} = \left( \frac{2\tilde{\zeta}}{3-\tilde{\zeta}} \right)^{2/3} -1
\end{equation}

\begin{figure}
\begin{center}
\includegraphics[width=10cm]{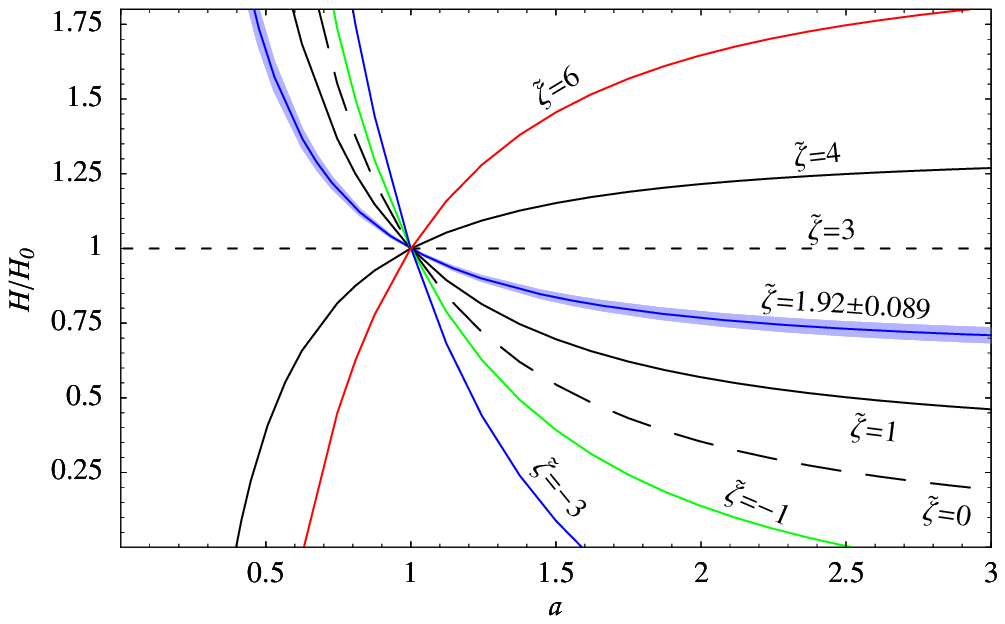}\\
\caption{Plot of $H(a,\tilde{\zeta})$ for different values of
$\tilde{\zeta}$ (see expression \eref{HubbleParameterAZ0}). $a$ is
the scale factor and $H_0$ is the Hubble constant. The long dashed
line corresponds to $\tilde{\zeta}=0$ (a flat matter-dominated
universe with null bulk viscosity) and the short dashed line
corresponds to $\tilde{\zeta}=3$ (the \emph{de Sitter} universe).
The blue line with the blue band correspond to a model with
$\tilde{\zeta}= 1.922\pm 0.089$. This is the best estimated value of
$\tilde{\zeta}$ coming from the SCP ``Union'' SNe Ia data set
analysis (see section \ref{SectionSNeTest}). The band corresponds to
the error at the 68.3\% confidence level.}
  \label{HubbleParameterAZ0plot}
\end{center}
\end{figure}

From the expression \eref{ScaleFactorTransition0leqZleq3Z0} we can
see that for values of $\tilde{\zeta}$ in the interval
$0<\tilde{\zeta}<1$ the transition between the decelerated epoch to
the accelerated one takes place in the future ($a_{\rm t}>1$). When
$\tilde{\zeta} \rightarrow 0$, the value of $a_{\rm t}$  tends to
infinity in the future. When $\tilde{\zeta}=1$ then the transition
takes place today ($a_{\rm t}=1$), when $1<\tilde{\zeta}<3$ the
transition takes place in the past of the universe ($0<a_{\rm t}<1$)
and when the value of $\tilde{\zeta}$ is closer to 3, the transition
is closer to the Big-Bang time (see figures
\ref{ScaleFactorDerivedadxPlot}--\ref{HubbleParameterAZ0plot} and
\ref{DecelerationParameterPlot}).

The curvature scalar $R$ is always positive and it has a very large
value when the scale factor is very small (at the Big-Bang time) and
it decreases forever tending to its minimum value $R=
\frac{4}{3}H_0^2 \tilde{\zeta}\;$ when $a \rightarrow \infty$ (see
section \ref{SectionScalarOfCurvatureZ0} and
\fref{PlotCurvatureScalar}) what is the de Sitter universe
(consistent with the expression
\eref{ScaleFactorProportionaldeSitter}). In the same way, the matter
density $\rho_{\rm m}$ is also very large when the scale factor is
very small (at the Big-Bang time) and it decreases forever tending
to its minimum value $\rho_{\rm m}= (H_0^2/ 24 \pi G)
\tilde{\zeta}^2$ when $a \rightarrow \infty \;$ (see section
\ref{SectionDensityMatterZ0}).

The possible value for $\tilde{\zeta}$ at least at the 99.9\%
confidence level estimated from the SCP ``Union'' SNe Ia data set is
in the interval $1<\tilde{\zeta}<3$. So, assuming the best estimated
value $\tilde{\zeta}=1.922 \pm 0.089$ (see
\tref{tableonlyMatterSummary_ZiHo}), the value of the scale factor
at the transition time between deceleration--acceleration epochs is
(see expression \eref{ScaleFactorTransition0leqZleq3Z0}): $a_{\rm
t}=0.42 \pm 0.03 \;$ ($z_{\rm t}=1.33 \pm 0.2)$. Note that the
transition comes directly from the model, it is not necessary to
assume any particular ansatz to estimate or induce it.

                    \subsubsection{Case $\tilde{\zeta} = 3$.}
                    \label{SectionZ0eq3}

In the expression \eref{ScaleFactorFor0z3} we put $\tilde{\zeta} = 3$ to obtain
\begin{equation} \label{ScaleFactorZ0equal3}
a(t) = \e ^{H_0(t-t_0)}
\end{equation}
This case corresponds to the \emph{de Sitter} universe. Figures
\ref{ScaleFactorFor0z3Plot}--\ref{ScaleFactorForZgeq3plot} show the behavior of expression
\eref{ScaleFactorZ0equal3} (the short-dashed lines).
The scale factor becomes zero just when $H_0(t-t_0) \rightarrow -\infty$ and when
$H_0(t-t_0) \rightarrow \infty$ the scale factor increases to infinity.

In this case the model predicts an universe in an \emph{eternal accelerated expansion}.
The curvature scalar and the matter density  are constants  with
values $R=12H_0^2$ and $\rho_{\rm m}= 3H_0^2/ 8 \pi G$ respectively. See sections
\ref{SectionScalarOfCurvatureZ0} and \ref{SectionDensityMatterZ0} for details.

Note that this model does not have a Big-Bang because the curvature scalar
and the matter density are regular for any value of the cosmic time.

                    \subsubsection{Case $\tilde{\zeta}>3$.}
                    \label{sectionScaleFactorZ0geq3}

This case predicts an universe \emph{expanding forever}, and this
expansion is \emph{always accelerated} (there is not any
decelerating epoch or acceleration-deceleration transition).

\begin{figure}
\begin{center}
\includegraphics[width=10cm]{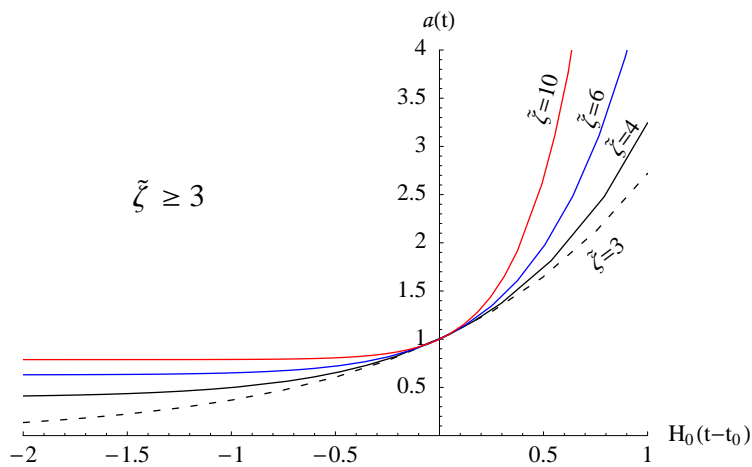}\\
\caption{Plot of $a(t,\tilde{\zeta})$ for different values of
$\tilde{\zeta}$ in the range $\tilde{\zeta} \geq 3$ (see expression
\eref{ScaleFactorFor0z3}). $H_0$ is the Hubble constant, $t$ is the
cosmic time and $t_0$ is its present value. The short dashed line
corresponds to $\tilde{\zeta}=3$ (the \emph{de Sitter} universe).}
\label{ScaleFactorForZgeq3plot}
\end{center}
\end{figure}

In this case when $t \rightarrow \infty$ then the scale factor tends
to the form of the de Sitter universe in the future (see expression
\eref{ScaleFactorProportionaldeSitter}) and when $t \rightarrow
-\infty$ then the universe tends to an \emph{Einstein static
universe} (defined by $\dot{a}, \ddot{a}=0$) where the value of the
scale factor is not zero but tends to its \emph{minimum} value:
\begin{equation}\label{ScaleFactorConditionForZgeq3}
\lim_{t \rightarrow -\infty} a(t) \equiv a^\ast =
\left(1-\frac{3}{\tilde{\zeta}}\right)^{2/3}
\end{equation}
and during the evolution of the universe the scale factor is an
increasing monotonic function with respect to the cosmic time. Thus,
for this case there was not a Big-Bang and the age of the universe
\emph{is not defined}. \Fref{ScaleFactorForZgeq3plot} shows some
plots of the expression \eref{ScaleFactorFor0z3} for different
values of $\tilde{\zeta}>3$.

When the scale factor has the value $a^\ast$  the curvature scalar
is zero and it is \emph{always increasing} from zero until to reach
its maximum value $R=\frac{4}{3}H_0^2 \tilde{\zeta}$ when $a
\rightarrow \infty$. See section \ref{SectionScalarOfCurvatureZ0}
and \fref{PlotCurvatureScalar} for details.

The matter density $\rho_{\rm m}$ has a similar behavior to the curvature scalar in the sense of
that when the scale factor has the value $a^\ast$ the matter density is zero and it is
\emph{always increasing} from zero until to reach its maximum value
$\rho_{\rm m}=(H_0^2/24 \pi G) \tilde{\zeta}^2$ when $a \rightarrow \infty$. See section
\ref{SectionDensityMatterZ0} and \fref{PlotDensityZ0} for details.

                  \subsubsection{Case $\tilde{\zeta}<0$.}
                  \label{SectionZ0leq0}

\begin{figure}
\begin{center}
\includegraphics[width=10cm]{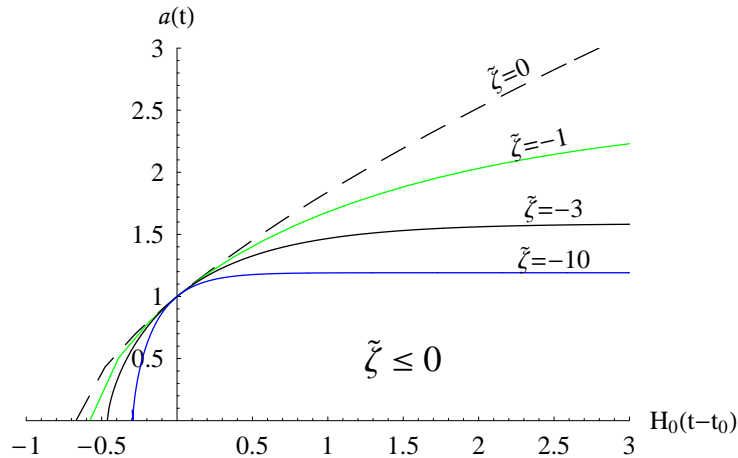}\\
\caption{Plot of $a(t,\tilde{\zeta})$ for different values of
$\tilde{\zeta}$ in the range $\tilde{\zeta} \leq 0$ (see expressions
\eref{ScaleFactorForZ0eq0Z0} and \eref{ScaleFactorFor0z3}). $H_0$ is
the Hubble constant, $t$ is the cosmic time and $t_0$ is its present
value. The long dashed line corresponds to $\tilde{\zeta}=0$ (a flat
matter-dominated universe with null bulk viscosity, see expression
\eref{ScaleFactorForZ0eq0Z0}).}
  \label{ScaleFactorForZleq0Z0plot}
\end{center}
\end{figure}

This case  predicts an \emph{eternal decelerated expanding} universe.
The universe begins with a Big-Bang and expands forever until to reach its \emph{maximum} value
$a^\ast = (1-3/\tilde{\zeta})^{2/3}$ when $t \rightarrow \infty$, becoming an Einstein static
universe in the future.
The Big-Bang took place in the cosmic time $t_{\rm B}$ defined by expression \eref{BigBangTime0Zo3}.
Figures \ref{ScaleFactorDerivedadxPlot}--\ref{HubbleParameterAZ0plot} and
\ref{ScaleFactorForZleq0Z0plot} show the behavior of the scale factor for this case.

In this case the curvature scalar has a \emph{transition} from positive to negative values at
$a_0 \equiv [(\tilde{\zeta}-3)/4\tilde{\zeta}]^{2/3}$. When $a \rightarrow 0$ then
$R\rightarrow \infty$ (the Big-Bang) and when $a=a^{\ast}$ the curvature scalar is zero.
The minimum value of the curvature scalar is \emph{negative} and it is reached at the value of the
scale factor  $\tilde{a} \equiv [(2-6/\tilde{\zeta})/5]^{2/3}$. See section
\ref{SectionScalarOfCurvatureZ0} and \fref{PlotCurvatureScalar} for details.

The matter density is always a decreasing function over its evolution starting from infinity
(when $a \rightarrow 0$) to zero (when the scale factor is $a^{\ast}$). See section
\ref{SectionDensityMatterZ0} and \fref{PlotDensityZ0} for details.

            \section{Cosmological parameters.}
            \label{SectionCosmologicalParameters}

            \subsection{Deceleration parameter $q$.}
            \label{SectionDecelerationParameterZ0}

We study the behavior of the \emph{deceleration parameter} $q$ function that is defined as
\begin{equation}\label{DecelerationParameterDefinition}
    q(a) \equiv  - \frac{\ddot{a}a}{\dot{a}^2}=-\frac{\ddot{a}}{a}
    \frac{1}{H^2}
\end{equation}
The term $\ddot{a}/a$ can be calculated from the second Friedmann equation, that for a
matter-dominated universe with bulk viscosity reads:
\begin{equation}\label{ScalarCurvature2dFriedmannEq}
\frac{\ddot{a}}{a} = -\frac{4 \pi G}{3} (\rho_{\rm m} - 9\zeta H)
\end{equation}
\noindent From the definition of $\tilde{\zeta}$ (see expression in \eref{DefinitionsZoBar}) we have
that
\begin{equation}\label{Definition2ZoBar}
\zeta=\left( \frac{H_0}{24\pi G} \right) \tilde{\zeta}
\end{equation}

\noindent On the other hand, from the first Friedmann equation
\eref{FriedmannEq1Matter} we have
\begin{equation}\label{Z0Densitymatter1}
\rho_{\rm m} = \frac{3}{8 \pi G}\; H^2(a)
\end{equation}

\noindent So, substituting the expressions \eref{Definition2ZoBar} and \eref{Z0Densitymatter1} in
\eref{ScalarCurvature2dFriedmannEq} we obtain:
\begin{equation}\label{2ndFriedmanEq_a_Z0}
\frac{\ddot{a}}{a}=\frac{1}{2} \left(\tilde{\zeta}  H_0 - H(a)  \right) H(a)
\end{equation}
Thus, substituting \eref{2ndFriedmanEq_a_Z0} in \eref{DecelerationParameterDefinition} yields
\begin{equation*}
    q(a)= \frac{1}{2} \left( 1- \tilde{\zeta} \frac{H_0}{H(a)} \right)
\end{equation*}
Using equation \eref{HubbleParameterAZ0} for $H(a)$ we arrive to

\begin{equation}\label{DecelerationParameteraZ0}
    q(a,\tilde{\zeta})= \frac{1}{2} \left[
 \frac{3-\tilde{\zeta} (1+2a^{3/2})}{3-\tilde{\zeta} (1-a^{3/2})}  \right]
\end{equation}

\begin{figure}
\begin{center}
\includegraphics[width=10cm]{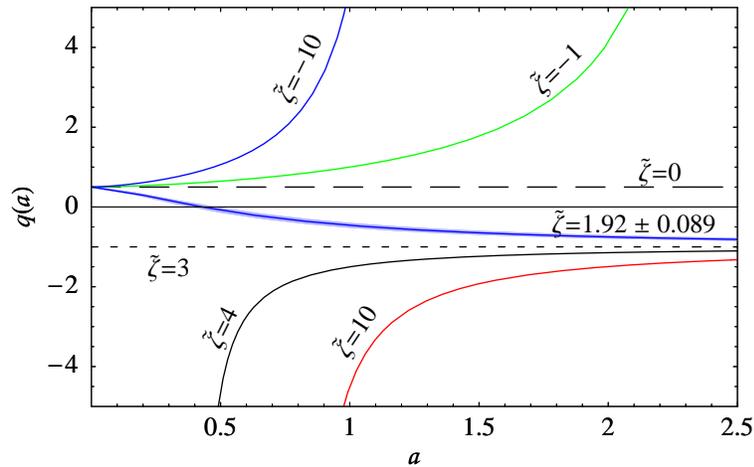}\\
\caption{Plot of deceleration parameter $q(a,\tilde{\zeta})$ for
different values of $\tilde{\zeta}$   (see expression
\eref{DecelerationParameteraZ0}). The long dashed line corresponds
to $\tilde{\zeta}=0$ (a flat matter-dominated universe with null
bulk viscosity). The short dashed line corresponds to
$\tilde{\zeta}=3$ (the \emph{de Sitter} universe). The blue line
with the blue band correspond to a model with $\tilde{\zeta}=
1.922\pm 0.089$. This is the best estimated value of $\tilde{\zeta}$
coming from the SCP ``Union'' SNe Ia data set analysis (see section
\ref{SectionSNeTest}). The band corresponds to the error at the
68.3\% confidence level.}
  \label{DecelerationParameterPlot}
\end{center}
\end{figure}

\noindent \Fref{DecelerationParameterPlot} shows the behavior of expression
\eref{DecelerationParameteraZ0}. Note that this expression does not depend on $H_0$. We summarize the
behavior of the deceleration parameter according to different values of $\tilde{\zeta}$ as
follows:

\begin{itemize}
  \item When $\tilde{\zeta} = 0$ we have $q= \frac{1}{2}$ that corresponds to a matter-dominated
universe with null bulk viscosity.

  \item When $\tilde{\zeta} = 3$ we have $ q= -1$ that corresponds to de Sitter universe.

  \item Case $0<\tilde{\zeta}<3$:
  It is always a decreasing function from $q(0)=\frac{1}{2}\;$ to $q(\infty)=-1$
  with a transition from positive to negative values in the value of the scale factor
  $a_{\rm t} \equiv [(3-\tilde{\zeta})/2\tilde{\zeta}]^{2/3}$ (see expression
  \eref{ScaleFactorTransition0leqZleq3Z0}).

  \item Case $\tilde{\zeta}>3$: When $a\rightarrow a^{\ast}$ then $q \rightarrow -\infty\;$ where
  $a^{\ast}= (1-3/\tilde{\zeta})^{2/3}$ is the \emph{minimum} value of the scale factor (see
expression \eref{ScaleFactorConditionForZgeq3}) and when $a \rightarrow \infty$ then $q \rightarrow
-1$. It is a negative and increasing function but it never becomes positive, its maximum value is -1.

  \item Case $\tilde{\zeta}<0$: When $a=0$ then $q= \frac{1}{2} \;$  and when $a \rightarrow
a^\ast$ then $q \rightarrow \infty$  where   $a^{\ast}=(1-3/\tilde{\zeta})^{2/3}$ is the
\emph{maximum} value of the scale factor. It is a positive and increasing monotonic function where
its minimum value is $\frac{1}{2}$.
\end{itemize}

From the expression \eref{DecelerationParameteraZ0} we obtain the deceleration parameter evaluated
\emph{today} as
\begin{equation}\label{DecelerationParametera1Z0}
    q(a=1,\tilde{\zeta}) = \frac{1-\tilde{\zeta}}{2}
\end{equation}

Note that when $\tilde{\zeta}=1$ the transition from the decelerated
to accelerated epochs of the universe takes place \emph{today}. For
$\tilde{\zeta} < 1$ we have a \emph{decelerated} universe in the
present and for $\tilde{\zeta} >1$ we have an \emph{accelerated} one
today.

Assuming the best estimated values for $\tilde{\zeta}$ from
\tref{tableonlyMatterSummary_ZiHo}, the deceleration parameter
\emph{today} is (see expression \eref{DecelerationParametera1Z0}):
$q_{\rm today}=-0.46 \pm 0.04 \;$.

            \subsection{The curvature scalar $R$.}
            \label{SectionScalarOfCurvatureZ0}

We calculate the curvature scalar $R$ in order to study the singularities of the model.  For a
flat universe it is defined as
\begin{equation}\label{ScalarCurvatureDefinition}
R = 6 \left[\frac{\ddot{a}}{a} + H^2\right]
\end{equation}

\noindent Using the expression \eref{2ndFriedmanEq_a_Z0} in \eref{ScalarCurvatureDefinition}
we obtain
\begin{equation}\label{Curvature2}
R=3 \left[\tilde{\zeta} H_0+3H(a) \right]H(a)
\end{equation}

\noindent Substituting the expression \eref{HubbleParameterAZ0} for $H(a)$ into \eref{Curvature2}
yields

\begin{equation}\label{ScalarCurvatureZ0}
R(a, \tilde{\zeta})= \frac{H_0^2}{3} \left[ \frac{(3-\tilde{\zeta})^2}{a^3} +
\frac{5\tilde{\zeta}(3-\tilde{\zeta})}{a^{3/2}} +
4\tilde{\zeta}^2 \right]
\end{equation}

\begin{figure}
\begin{center}
\includegraphics[width=13cm]{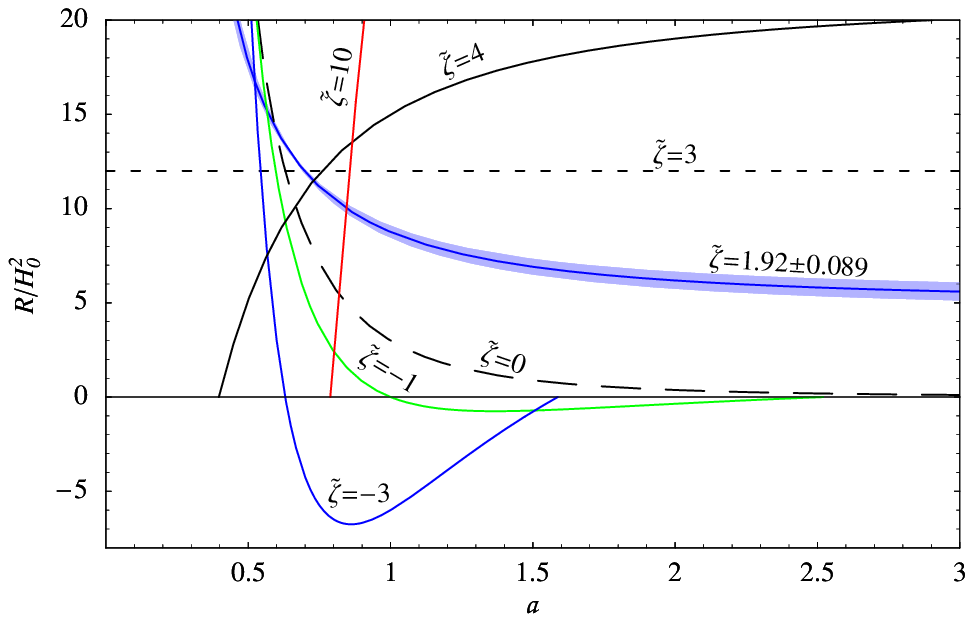}\\
\caption{Plot of the curvature scalar $R(a,\tilde{\zeta})$ as a
function of the scale factor for different values of $\tilde{\zeta}$
(see expression \eref{ScalarCurvatureZ0}). The long dashed line
corresponds to $\tilde{\zeta}=0$ (a flat matter-dominated universe
with null bulk viscosity). The short dashed line corresponds to
$\tilde{\zeta}=3$ (the \emph{de Sitter} universe). The blue line
with the blue band correspond to a model with $\tilde{\zeta}=
1.922\pm 0.089$. This is the best estimated value of $\tilde{\zeta}$
coming from the SCP ``Union'' SNe Ia data set analysis (see section
\ref{SectionSNeTest}). The band corresponds to the error at the
68.3\% confidence level.}
  \label{PlotCurvatureScalar}
\end{center}
\end{figure}

\noindent \Fref{PlotCurvatureScalar} shows some plots of expression \eref{ScalarCurvatureZ0}.
From this expression we note that:

\begin{itemize}
\item When $\tilde{\zeta}=0$ then $ R= 3 H_0^2 / a^3$ that corresponds to a matter-dominated
universe with null bulk viscosity.

  \item When $\tilde{\zeta} = 3$ then $R= 12 H_0^2$ that corresponds to de Sitter universe.

  \item Case $0<\tilde{\zeta}<3$:  When $a \rightarrow 0$ then $R\rightarrow \infty$ (the Big-Bang
singularity). It is \emph{always} a positive and decreasing function until to reach its minimum
value $R=\frac{4}{3}H_0^2 \tilde{\zeta}$ when $a \rightarrow \infty$.

  \item Case  $\tilde{\zeta}>3$:
It is \emph{always} a positive and increasing function from zero at $a=a^{\ast}$ (where
$a^{\ast}= (1-3/\tilde{\zeta})^{2/3}$ is the minimum value of the scale factor) until to reach its
maximum value $R= \frac{4}{3}H_0^2 \tilde{\zeta}$ when $a \rightarrow \infty$.

  \item Case $\tilde{\zeta}<0$:  When $a \rightarrow 0$ then $R\rightarrow \infty$ (the Big-Bang
singularity). The curvature scalar is zero at the values of the scale factor $a_0$ and $a^{\ast}$,
where
\begin{equation}
a_0=\left( \frac{\tilde{\zeta}-3}{4\tilde{\zeta}} \right)^{2/3}
\end{equation}
and $a^{\ast}$ is the \emph{maximum} value for the scale factor defined in expression
\eref{ScaleFactorConditionForZgeq3}. At $a_0$ there is a \emph{transition} from positive to negative
values of $R$. When $0<a<a_0\;$ the curvature scalar is a positive and decreasing function. For
$a_0<a<a^{\ast}$ the value of $R$ is \emph{negative}. The minimum value of the curvature scalar is
$R = -\frac{3}{4}H_0^2\tilde{\zeta}^2$ at the value of the scale factor $\tilde{a}$
\begin{equation}
\tilde{a}=\left[ \frac{2}{5} \left(\frac{\tilde{\zeta}-3}{\tilde{\zeta}} \right) \right]^{2/3}
\end{equation}

\end{itemize}

            \subsection{The matter density $\rho_{\rm m}$.}
            \label{SectionDensityMatterZ0}

\begin{figure}
\begin{center}
\includegraphics[width=12cm]{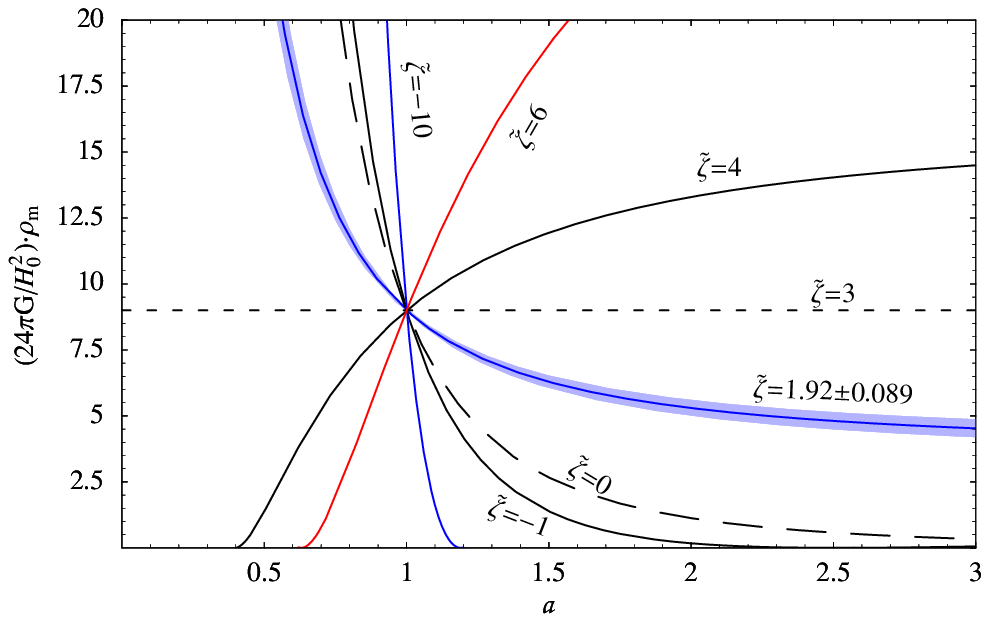}\\
\caption{Plot of the matter density $\rho_{\rm m}(a,\tilde{\zeta})$
as a function of the scale factor for different values of
$\tilde{\zeta}$ (see expression \eref{Z0DensitymatterA}). The long
dashed line corresponds to $\tilde{\zeta}=0$ (a flat
matter-dominated universe with null bulk viscosity). The short
dashed line corresponds to $\tilde{\zeta}=3$ (the \emph{de Sitter}
universe). The blue line with the blue band correspond to a model
with $\tilde{\zeta}= 1.922\pm 0.089$. This is the best estimated
value of $\tilde{\zeta}$ coming from the SCP ``Union'' SNe Ia data
set analysis (see section \ref{SectionSNeTest}). The band
corresponds to the error at the 68.3\% confidence level.}
  \label{PlotDensityZ0}
\end{center}
\end{figure}

We calculate the matter density $\rho_{\rm m}$ predicted for this model.
We substitute the Hubble parameter $H(a)$ (expression \eref{HubbleParameterAZ0}) into the expression
\eref{Z0Densitymatter1} for the matter density, yielding

\begin{equation}\label{Z0DensitymatterA}
\rho_{\rm m}(a, \tilde{\zeta}) = \frac{H_0^2}{24 \pi G} \left[ \tilde{\zeta} +
\frac{3-\tilde{\zeta}}{a^{3/2}} \right]^2
\end{equation}
\Fref{PlotDensityZ0} shows some plots of expression \eref{Z0DensitymatterA}. From this expression we
note that:

\begin{itemize}
  \item When $\tilde{\zeta} = 0$ then $\rho_{\rm m} = (3H_0^2/8\pi G) \, a^{-3}$ that corresponds
to a matter-dominated universe with null bulk viscosity.

  \item When $\tilde{\zeta} = 3 $ then $ \rho_{\rm m} = 3H_0^2/8\pi G$ that corresponds to de Sitter
universe.

  \item Case $0<\tilde{\zeta}<3$: When $a \rightarrow 0$ then $\rho_{\rm m} \rightarrow \infty$ (the
  Big-Bang singularity). It is a decreasing function until to reach its \emph{minimum} value
$\rho_{\rm m} =(H^2_0/24\pi G)\, \tilde{\zeta}^2$ when $a \rightarrow \infty$.

  \item Case $\tilde{\zeta}>3$: When $a=a^{\ast}$ then $\rho_{\rm m}(a^\ast)=0$ where
$a^\ast=(1-3/\tilde{\zeta})^{2/3}$ is the \emph{minimum} value of the scale factor (see expression
\eref{ScaleFactorConditionForZgeq3} and section \ref{sectionScaleFactorZ0geq3}).
It is an increasing monotonic function from zero until to reach its \emph{maximum} value of
  $\rho_{\rm m} =(H^2_0/24\pi G)\,\tilde{\zeta}^2$ when $a \rightarrow \infty$.
In this case, the matter density increases as in the Hoyle's steady state cosmology
\cite{Hoyle1958}.

  \item Case $\tilde{\zeta}<0$: When $a \rightarrow 0$ then $\rho_{\rm m} \rightarrow \infty$ (the
Big-Bang singularity). It is a decreasing function until to reach its \emph{minimum} value in
$\rho_{\rm m}(a^\ast)=0$.
\end{itemize}

            \subsection{The age of the universe.}
            \label{SectionAgeUniverseZ0}

We define the \emph{age} of the universe as the elapsed time between the Big-Bang time $t_{\rm B}$
until today $t_0$. So, from the expression \eref{BigBangTimeZ0eq0} for the case $\tilde{\zeta}=0$ we
 have
\begin{equation}\label{AgeUniverseZequal0Z0}
{\mbox {Age}} \equiv |t_{\rm B}-t_0|= \frac{2}{3 H_0}, \qquad {\mbox {for}} \quad \tilde{\zeta}=0
\end{equation}
And from the expression \eref{BigBangTime0Zo3} for the case $\tilde{\zeta} < 3$
\begin{equation}\label{AgeUniverse0z3Z0}
{\mbox {Age}} \equiv |t_{\rm B}-t_0|= -\frac{2}{\tilde{\zeta} H_0} \ln\left( 1- \frac{\tilde{\zeta}}{3}
\right) , \qquad {\mbox {for}} \quad \tilde{\zeta}<3
\end{equation}

\begin{figure}
\begin{center}
\includegraphics[width=8cm]{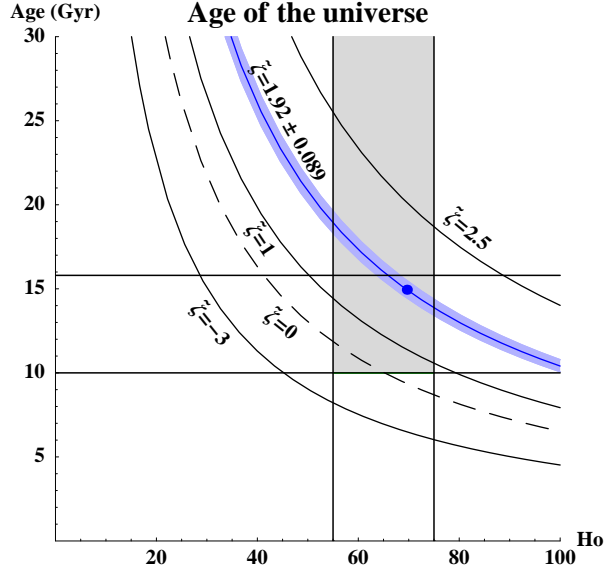}\\
\caption{Plot of the age of the universe in units of Gigayears (Gyr)
with respect to $H_0$ in units of  $({{\rm km}}/{{\rm s}}){{\rm
Mpc}}^{-1}$, for different values of $\tilde{\zeta}$ (see expression
\eref{AgeUniverseZequal0Z0}, \eref{AgeUniverse0z3Z0}). The blue
point is located at $14.957 \pm 0.422$ Gyr that corresponds to the
best estimated value for the age of the universe coming from the SCP
``Union'' SNe Ia data set (where $H_0=69.62 \pm 0.59 ({{\rm
km}}/{{\rm s}}){{\rm Mpc}}^{-1}$, see section \ref{SectionSNeTest}).
The blue line with the blue band correspond to a model evaluated at
the best estimated value for $\tilde{\zeta}= 1.922\pm 0.089$. The
vertical lines correspond to the interval $H_0=[55,75]\,({{\rm
km}}/{{\rm s}}) {{\rm Mpc}}^{-1}$, it is the permitted region
according to values of $H_0$ consistent with the distance moduli
used to derive ages for Galactic globular clusters from the
\emph{Hipparcos} parallaxes (see ref. \cite{CarretaEtal2000}). The
horizontal lines corresponds to the constraint for the age of the
universe from the oldest globular clusters (Age$=12.9 \pm 2.9$ Gyr
\cite{CarretaEtal2000}). So, the shaded area is the consistent
region for the age of the universe. The dashed line corresponds to
$\tilde{\zeta}=0$, a flat matter-dominated universe with null bulk
viscosity.}
  \label{plotAgesUniverseHoversusAgeShadowZ0}
\end{center}
\end{figure}

Note that for the case $\tilde{\zeta} \geq 3$ the age of the
universe \emph{is not defined}. On the other hand, evaluating the
expression \eref{AgeUniverse0z3Z0} in the best estimated values of
$\tilde{\zeta}$ and $H_0$ from the SCP ``Union'' SNe Ia data set
(see \tref{tableonlyMatterSummary_ZiHo}) yields an age of $14.957
\pm 0.422$ Gyr. The errors are at 68.3\% confidence level. We found
that this age is in perfect agreement with the constraints on the
age of the universe coming from the oldest globular clusters
\cite{CarretaEtal2000}. Figure
\ref{plotAgesUniverseHoversusAgeShadowZ0} shows the age of the
universe for different values of $\tilde{\zeta}$ and the best
estimate, the vertical lines correspond to $H_0=[55,75]$ $({{\rm
km}}/{{\rm s}}){{\rm Mpc}}^{-1}$ \footnote{Hereinafter the units of
$H_0$ are expressed in $({{\rm km}}/{{\rm s}}){{\rm Mpc}}^{-1}$}, it
is the permitted region according to values of $H_0$ consistent with
the distance moduli used to derive ages for Galactic globular
clusters from the \emph{Hipparcos} parallaxes. As a reference, the
predicted age of the universe for $\Lambda$CDM (Lambda Cold Dark
Matter) cosmological model with a flat universe and using the best
estimated values of $\Omega_{{\rm m}0}$ and $H_0$ shown in
\tref{tableonlyMatterSummary_ZiHo} is $13.750\pm 0.29$ Gyr.

        \section{Thermodynamics and the local entropy.}
        \label{Sectionthermodynamics}

The law of generation of the \textit{local} entropy in the FRW space--time is found to be
\cite{Weinberg,Misner,Hofmann}
\begin{equation}\label{entropy_definition}
T \, \nabla_{\nu} s^{\nu} = \zeta \nabla_{\nu} u^{\nu} = 3H \zeta
\end{equation}
\noindent where $T$ is the temperature and $\nabla_{\nu} s^{\nu} $ is the rate at which entropy is
being generated in an unit volume. Then, the second law of the thermodynamics can be written as
\begin{equation}\label{2ndLawThermodynamics}
 T \nabla_{\nu} s^{\nu} \geq 0
\end{equation}
\noindent which, from the expression \eref{entropy_definition}, it implies that $ 3H\zeta \geq 0$.

Since the Hubble parameter $H$ is positive in an expanding universe then $\zeta$ has
to be positive in order to preserve the validity of the second law of the
thermodynamics. Thus, equation \eref{2ndLawThermodynamics} can be written for this model as
\begin{equation}\label{entropy_condition}
\tilde{\zeta} \geq 0
\end{equation}

                \section{Type Ia Supernovae test.}
                \label{SectionSNeTest}

We analyze and constrain the viability of the model using the type
Ia supernovae  (SNe Ia) observations and the entropy tests (see
expression \eref{entropy_condition} and
\sref{Sectionthermodynamics}). For that, we calculate the
\textit{best estimated values} for the parameters $\tilde{\zeta}$
and $H_0$  and the \emph{goodness-of-fit} of the model to the data
by $\chi^2$-minimization and then compute the confidence intervals
for $(\tilde{\zeta}, H_0)$  and the probability density functions
for $\tilde{\zeta}$ (marginalizing over $H_0$) to constrain their
possible values.

We perform the statistical analysis using the ``Union'' SNe Ia data
set of ``The Supernova Cosmology Project'' (SCP) \cite{Kowalski2008}
composed by 307 type Ia supernovae brought together from 13
independent data sets.

We use the definition of luminosity distance $d_L$
\cite{Riess2004,Riess2006}, \cite{TurnerRiess}--\cite{Copeland} in a
flat cosmology, it is
\begin{eqnarray}
    d_L(z, \tilde{\zeta}, H_0) &=& c(1+z) \int_0^z \frac{dz'}{H(z', \tilde{\zeta}, H_0)}
    \label{luminosity_distance1}
\end{eqnarray}

\noindent where $H(z, \tilde{\zeta}, H_0)$ is the Hubble parameter
(expression \eref{SolutionFinalToODE-Z0})  and `$c$' is the speed of
light. The \emph{theoretical distance moduli} for the $k$-th
supernova with redshift $z_k$ is defined as
\begin{equation}\label{distanceModuli}
 \mu^{{\rm t}}(z_k, \tilde{\zeta}, H_0) \equiv m-M=5\log_{10} \left[
 \frac{d_L(z_k, \tilde{\zeta}, H_0)}{{\rm Mpc}} \right] +25
\end{equation}
\noindent where $m$ and $M$ are the apparent and absolute magnitudes of the SNe Ia
respectively and the superscript `t' stands for \textit{theoretical}.
We construct the statistical $\chi^2$ function as

\begin{equation}\label{ChiSquareDefinition}
\chi^2 (\tilde{\zeta}, H_0)
   \equiv \sum_{k = 1}^n
   \frac{\left[\mu^{{\rm t}} (z_k, \tilde{\zeta}, H_0) - \mu_k \right]^2}{\sigma_k^2}
\end{equation}

\noindent where $\mu_k$ is the \emph{observational} distance moduli for the $k$-th supernova,
$\sigma_k^2$ is the variance of the measurement and $n$ is the amount of supernovae in the data set.

Once constructed the $\chi^2$ function \eref{ChiSquareDefinition},
we  numerically minimize it to compute the ``\textit{best
estimates}'' for the free parameters of the model (i.e.,
$\tilde{\zeta}, H_0$). The  $\chi^2$ function measures the
goodness-of-fit of the model to data. The \textit{probability
density function} (\textbf{pdf}) is defined as
\begin{equation}
\mbox{pdf}(\tilde{\zeta}, H_0)= cte \cdot {{\rm e}}^{-\chi^2/2}
\end{equation}
where `\emph{cte}' is a normalization constant.

\begin{figure}
\begin{center}
\includegraphics[width=12cm]{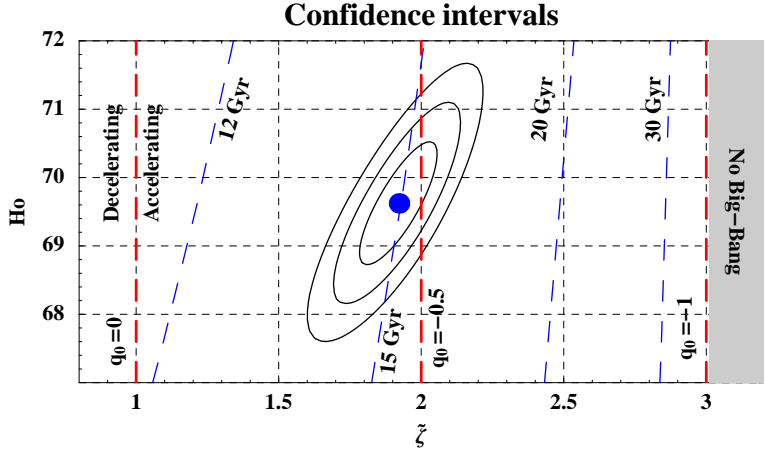}\\
\caption{Confidence intervals for $\tilde{\zeta}$ and  $H_0$ of a
bulk viscous matter-dominated universe model with constant bulk
viscosity. The best estimated values and confidence intervals  were
computed using the SCP ``Union'' SNe Ia data set where it is found
$\tilde{\zeta}=1.922 \pm 0.089$ and $H_0=69.62 \pm 0.59 \,({{\rm
km}}/{{\rm s}}) {{\rm Mpc}}^{-1}$ as best estimates (see
\tref{tableonlyMatterSummary_ZiHo}). The shown confidence intervals
correspond to $68.3\%$, $95.4\%$ and $99.73\%$ of probability for
the estimated value. $H_0$ is in units of $\,({{\rm km}}/{{\rm s}})
{{\rm Mpc}}^{-1}$ and $\tilde{\zeta}$ is dimensionless.  Note that
at least at the $99.73 \%$ confidence level the possible values for
$\tilde{\zeta}$ are inside of the interval $1<\tilde{\zeta}<3$
indicating an accelerating universe {\it today} with high confidence
level (see subsection \ref{Section0Z03}). The blue dashed lines show
different ages of the universe according to the values of
$\tilde{\zeta}$ and $H_0$ (see expression \eref{AgeUniverse0z3Z0}).
The red vertical dashed lines indicate values of the deceleration
parameter evaluated \emph{today} (see expression
\eref{DecelerationParametera1Z0}) being the line $q_0=0$ the limit
between a decelerated--accelerated universe \emph{today}. The
estimated age of the universe is $14.957 \pm 0.422$ Gyr, where the
error is at 68.3\% confidence level. We found that this age is in
perfect agreement with the constraints on the age of the universe
coming from the oldest globular clusters \cite{CarretaEtal2000} (see
\fref{plotAgesUniverseHoversusAgeShadowZ0}). The models that are in
the grey region ($\tilde{\zeta} \geq3$) do not have a Big-Bang in
the past of the universe as the label indicates.}
\label{ConfInterViscousMatterGoldEssenceZ0HoOver}
\end{center}
\end{figure}

So, we minimize the $\chi^2$ function to obtain the results shown in
\tref{tableonlyMatterSummary_ZiHo}. The confidence intervals for
$\tilde{\zeta}$ and $H_0$ are shown in
\fref{ConfInterViscousMatterGoldEssenceZ0HoOver}. Note that at least
at the 99.73\% confidence level the possible values for
$\tilde{\zeta}$ are inside of the interval $1<\tilde{\zeta}<3$. The
models in this interval indicate the existence of a Big-Bang in the
past of the universe, they have an accelerated expansion today and
predict an age of the universe $\sim 13-18$ Gyr at the 99.73 \%
confidence level (see
\fref{ConfInterViscousMatterGoldEssenceZ0HoOver} and subsections
\ref{Section0Z03}, \ref{SectionAgeUniverseZ0}).

In order to compare the results shown in
\tref{tableonlyMatterSummary_ZiHo} (in particular the measurement of
the goodness-of-fit to data $\chi^2_{\rm{d.o.f.}}$ and the best
estimated value for $H_0$) with other models we compute also the
best estimated values for the $\Lambda$CDM model using the same SNe
Ia data set. For the $\Lambda$CDM model, we assume a flat cosmology
with $H_0$ and $\Omega_{{\rm m}0}$ as free parameters (implying that
the cosmological constant density parameter is $\Omega_{\Lambda
0}=1-\Omega_{{\rm m}0}$). The results are shown in
\Tref{tableonlyMatterSummary_ZiHo}.

We find that the best estimated values for $H_0$ and the
goodness-of-fit to data $\chi^2_{\rm{d.o.f.}}$ from $\Lambda$CDM
model are very similar to those obtained from the bulk viscous model
using the same SCP ``Union'' SNe Ia data set and in agreement with
that reported by 5 year WMAP ($H_0=70.5 \pm 1.3$ (km/s)Mp${\rm
c}^{-1}$, see \cite{WMAP}).

\begin{table}
  \centering
\begin{tabular}{c | c c c | c c}
 \multicolumn{6}{c}{\textbf{Viscous model $\zeta=$ constant}} \\

\hline

Model & $H_0$ &$\tilde{\zeta}$ & $\Omega_{{\rm m}0}$ & $\chi^2_{{\rm
min}}$ &
$\chi^2_{{\rm d.o.f.}}$ \\
\hline \hline

Constant bulk viscosity & $69.62 \pm 0.59$ & $1.922 \pm 0.089$ & 1 &
314.57 & 1.031 \\
$\Lambda$CDM & $70.01 \pm 0.59$ & --- &$0.278 \pm 0.027$ &
311.84 & 1.022 \\
\hline

\end{tabular}
\caption{Summary of the best estimates of $\tilde{\zeta}$ and $H_0$
for the bulk viscous matter-dominated universe model with a constant
bulk viscosity coefficient. Also, it is shown the best estimates for
the $\Lambda$CDM model in order to compare the values of $H_0$ and
$\chi^2_{{\rm min}}$ of both models. It is found that the best
estimates for $H_0$ and their corresponding $\chi^2_{{\rm min}}$
computed for both models and using the same SNe Ia data set are
almost the same value and in agreement with the reported value by 5
year WMAP ($H_0=70.5 \pm 1.3$ (km/s)Mp${\rm c}^{-1}$, see
\cite{WMAP}). For the $\Lambda$CDM model, we assumed a flat
cosmology with $H_0$ and $\Omega_{{\rm m}0}$ as free parameters,
where $\Omega_{{\rm m}}$ is the matter density parameter. Note that
in the constant bulk viscous model the value of $\Omega_{{\rm m}}$
is not estimated but it is assumed to have the value of one as part
of the model. The best estimates were computed by a Bayesian
statistical analysis using the SCP ``Union'' 2008 compilation data
set composed by 307 SNe Ia from 13 independent data sets
\cite{Kowalski2008}. $H_0$ is in units of $({{\rm km}}/{{\rm
s}}){{\rm Mpc}}^{-1}$ and $\tilde{\zeta}$, $\Omega_{{\rm m}}$ are
dimensionless. The subscript ``d.o.f.'' stands for \emph{degrees of
freedom} and the errors are at 68.3\% confidence level.
\Fref{ConfInterViscousMatterGoldEssenceZ0HoOver} shows the
confidence intervals.}
  \label{tableonlyMatterSummary_ZiHo}
\end{table}

            \section{Marginalization over $H_0$}
            \label{SectionMarginalizationHo}

Now, we construct a {\bf pdf} for the bulk viscous coefficient
$\tilde{\zeta}$ marginalizing over the Hubble constant $H_0$ in
order to have to $\tilde{\zeta}$ as the only free parameter of the
model. We use three different priors to marginalize $H_0$:

\begin{enumerate}
  \item \emph{Constant} prior over $H_0$ (see \ref{SectionAnalyticalMargMethod}).

  \item  \emph{Gaussian} prior over $H_0$ centered at $H_0=70.5 \pm 1.3\,({{\rm km}}/{{\rm s}})
{{\rm Mpc}}^{-1}$ and $H_0=72 \pm 8 \,({{\rm km}}/{{\rm s}}) {{\rm Mpc}}^{-1}$ as reported by 5 year
 WMAP data  \cite{WMAP} and HST Cepheid variable star observations \cite{Freedman2001} respectively
(see \ref{SectionNumericalGaussMargMethod}).

  \item  \emph{Dirac delta} prior over $H_0$ centered at  $H_0=70.5\,({{\rm km}}/{{\rm s}})
{{\rm Mpc}}^{-1}$ and $H_0=72 \,({{\rm km}}/{{\rm s}}) {{\rm Mpc}}^{-1}$ reported as mentioned
above (see \ref{SectionDiracDeltaPriorMargMethod}).
\end{enumerate}

 \begin{table}
  \centering
 \begin{tabular}{c|c c c}
  \multicolumn{4}{c}{\textbf{Viscous model $\zeta=$ constant}}\\
  \multicolumn{4}{c}{Best estimates for $\tilde{\zeta}$}\\
   \hline
   \emph{Marg.} &  $\tilde{\zeta}$&$\chi^2_{{\rm min}}$&$\chi^2_{{\rm d.o.f.}}$ \\
  \hline \hline
  (i) & $1.922 \pm 0.089$ & 331.30 & 1.08  \\
  (ii) & $1.941 \pm 0.084 $ & 315.26 & 1.03 \\
  (iii) & $1.924 \pm 0.089$ & 314.91 & 1.02 \\
  (iv) & $2.026 \pm 0.051$ & 316.75 & 1.03 \\
  (v) & $2.195 \pm 0.049$ & 330.41 & 1.07 \\
  \hline
 \end{tabular}
\caption{Summary of the best estimates of $\tilde{\zeta}$ for the
bulk viscous matter-dominated universe model with a constant bulk
viscosity coefficient. The best estimates were computed by a
Bayesian statistical analysis using the SCP ``Union'' 2008
compilation data set composed by 307 SNe Ia from 13 independent data
sets \cite{Kowalski2008}. The errors are at 68.3\% confidence level
and $\tilde{\zeta}$ is dimensionless. \Fref{pdfViscosity0} shows the
probability density functions for $\tilde{\zeta}$. We have
marginalized over the Hubble constant $H_0$ using three different
priors described in section \ref{SectionMarginalizationHo}.
 In the first column of the table the label ``\emph{Marg.}''  indicates the marginalization method
 over $H_0$ used to obtain the results. The numbers correspond to:\\

\begin{enumerate}
  \item \emph{Constant} prior over $H_0$ (see \ref{SectionAnalyticalMargMethod}).
  \item  \emph{Gaussian} prior over $H_0$ centered at
 $H_0=70.5 \pm 1.3\,({{\rm km}}/{{\rm s}}) {{\rm Mpc}}^{-1}$, as reported by 5 year WMAP data
 \cite{WMAP} (see \ref{SectionNumericalGaussMargMethod}).
  \item \emph{Gaussian} prior over $H_0$  centered at
 $H_0=72 \pm 8 \,({{\rm km}}/{{\rm s}}) {{\rm Mpc}}^{-1}$, as reported by HST Cepheid
 variable star observations \cite{Freedman2001}  (see \ref{SectionNumericalGaussMargMethod}).
  \item  \emph{Dirac delta} prior over $H_0$ located at
 $H_0=70.5 \,({{\rm km}}/{{\rm s}}) {{\rm Mpc}}^{-1}$, as reported by 5 year WMAP data
 (see \ref{SectionDiracDeltaPriorMargMethod}).
  \item \emph{Dirac delta} prior over $H_0$  located at
 $H_0=72 \,({{\rm km}}/{{\rm s}}) {{\rm Mpc}}^{-1}$, as reported by HST Cepheid
 variable star observations  (see \ref{SectionDiracDeltaPriorMargMethod}).
\end{enumerate} }
  \label{tableonlyMatterSummaryZ012}
 \end{table}

      \subsection{Marginalization assuming a constant prior over $H_0$.}

We marginalize over $H_0$ by assuming a \textit{constant} prior  as
described in \ref{SectionAnalyticalMargMethod}. We use the
$\chi^2_{{\rm cp}}$ function \eref{chi_AnalyticalIntegration}
instead of $\chi^2$ function \eref{ChiSquareDefinition} to perform
the  supernova analysis. The results are:

\begin{itemize}
\item $\tilde{\zeta}= 1.922 \pm 0.089, \;$ with a $\;
\chi^2_{\rm{min}}= 331.30 \;$ ($\chi^2_{\rm{d.o.f.}}=1.08$).
\end{itemize}
The error is at 68.3\% confidence level. We plot the \textbf{pdf} of
$\tilde{\zeta}$ to show the value that maximizes such a function,
i.e., the best estimated value of $\tilde{\zeta}$. This result is
shown in \tref{tableonlyMatterSummaryZ012} and \fref{pdfViscosity0}.

            \subsection{Marginalization assuming a Gaussian prior over $H_0$.}

We marginalize over $H_0$ by assuming a \emph{Gaussian} prior as described in
\ref{SectionNumericalGaussMargMethod}.
We use two different central values for the Gaussian prior for $H_0$, they are
$H_0=70.5 \pm 1.3\,({{\rm km}}/{{\rm s}}){{\rm Mpc}}^{-1}$ and
$H_0=72 \pm 8 \,({{\rm km}}/{{\rm s}}) {{\rm Mpc}}^{-1}$ as suggested by 5 year WMAP data
\cite{WMAP} and HST Cepheid variable star observations \cite{Freedman2001} respectively. See section
\ref{SectionNumericalGaussMargMethod} for details.
The results are shown in \tref{tableonlyMatterSummaryZ012} and  \fref{pdfViscosity0}.

            \subsection{Marginalization assuming a Dirac delta prior over $H_0$.}

\begin{figure}
\begin{center}
  \includegraphics[width=5cm]{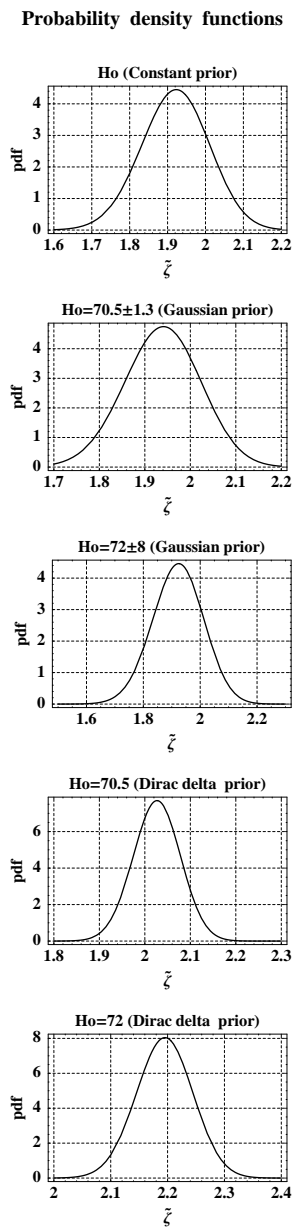}\\
  \caption{Probability distribution functions for the coefficient $\tilde{\zeta}$ in a bulk viscous
matter-dominated  universe model with a constant bulk viscosity. We
have marginalized over the Hubble constant $H_0$ using three
different priors described in section
\ref{SectionMarginalizationHo}. The best estimates and pdf's were
computed by a Bayesian statistical analysis using the SCP ``Union''
2008 compilation data set composed by 307 SNe Ia from 13 independent
data sets \cite{Kowalski2008}. \Tref{tableonlyMatterSummaryZ012}
summarizes the best estimated values of $\tilde{\zeta}$ using the
different marginalizations. The first figure (on top) corresponds to
the marginalization using a \emph{constant} prior over $H_0$ (see
\ref{SectionAnalyticalMargMethod}). The second and third figures
(from top to bottom) correspond to the marginalization using a
\emph{Gaussian} prior over $H_0$ centered at $H_0=70.5 \pm
1.3\,({{\rm km}}/{{\rm s}}) {{\rm Mpc}}^{-1}$ and $H_0=72 \pm 8
\,({{\rm km}}/{{\rm s}}) {{\rm Mpc}}^{-1}$ as reported by 5 year
WMAP  data \cite{WMAP} and HST Cepheid variable star observations
\cite{Freedman2001} respectively. The fourth and fifth figures
correspond to the marginalization using a \emph{Dirac delta} prior
over $H_0$ located at $H_0=70.5\,({{\rm km}}/{{\rm s}}) {{\rm
Mpc}}^{-1}$ and $H_0=72 \,({{\rm km}}/{{\rm s}}) {{\rm Mpc}}^{-1}$,
as mentioned above.}
  \label{pdfViscosity0}
\end{center}
\end{figure}

Finally, we marginalize over $H_0$ by assuming a \emph{Dirac delta} prior as described in
\ref{SectionDiracDeltaPriorMargMethod}.
In practice, it means to assume a specific value for $H_0$. In the same way as above we use two
different values for $H_0$, they are $H_0=70.5\,({{\rm km}}/{{\rm s}}){{\rm Mpc}}^{-1}$ and
$H_0=72\,({{\rm km}}/{{\rm s}}){{\rm Mpc}}^{-1}$. The results are shown in
\tref{tableonlyMatterSummaryZ012} and  \fref{pdfViscosity0}.

        \section{Conclusions.}
        \label{SectionConclusions}

We performed a detailed study of a bulk viscous matter-dominated
universe model with constant bulk viscous coefficient $\zeta$. The
only component of this model is a pressureless fluid with an
\emph{assumed} constant bulk viscosity  driving the present
accelerated expansion of the universe. We analyzed the different
possible scenarios for the behavior and evolution of the universe
predicted by the model according to the value of the dimensionless
bulk viscous coefficient $\tilde{\zeta}$. In general, the model
predicts an \emph{expanding} universe for any value of
$\tilde{\zeta}$ and we find that for values of  $\tilde{\zeta}$ in
the range  $0<\tilde{\zeta}<3$ the model predicts an universe with a
Big-Bang in the past and then starting with a \emph{decelerated}
expansion epoch followed by a \emph{transition} to an accelerated
epoch in late times.

The case $\tilde{\zeta}=3$ corresponds to the de Sitter universe.
When $\tilde{\zeta}>3$ the universe is always in an accelerated
expansion and the value of the scale factor tends to infinity in the
future (in a similar way than the de Sitter universe) and where the
curvature scalar goes to a positive constant value. In this case
there is not a Big-Bang nor a defined age nor origin of the universe
because the scale factor has a minimum value greater than zero  in
the past of the universe (when $t \rightarrow -\infty$), where also,
the curvature scalar and the matter density are zero, i.e., we have
the Einstein static universe in the past.

When $\tilde{\zeta}<0$ the universe is always in a decelerated
expansion and the scale factor tends to a finite value in the future
(it tends to the Einstein static universe in the future) where also,
the curvature scalar and the matter density go to zero. For
$\tilde{\zeta}$ in this range there is always a Big-Bang in the
past.

For any value of $\tilde{\zeta}$ there \emph{does not} exist any
recollapse epoch or transition between expansion to contraction
epochs. When $\tilde{\zeta}=1$ the transition between a decelerated
to accelerated epoch takes place \emph{today} (i.e., $q_0=0$ for
$\tilde{\zeta}=1$ ). For $0<\tilde{\zeta} < 1$ the transition
between the deceleration--acceleration epoch takes place in the
future and for $1<\tilde{\zeta} < 3$ the transition took place in
the past.

On the other hand, we estimate the values of $\tilde{\zeta}$ and
$H_0$ using the SCP Union SNe Ia data set where it is found that the
estimated values of $\tilde{\zeta}$ are positives, in agreement with
the second law of thermodynamics. We note that the best estimated
value for $H_0$ is in agreement with that reported by 5 year WMAP
and estimated from $\Lambda$CDM model (using the same SNe Ia data
set) and the minimum value of the $\chi^2$ function ($\chi^2_{\rm
d.o.f.}$) obtained is of the same order than that obtained for
$\Lambda$CDM model indicating that this is a competitive model to
fit the supernova observations.

We estimate and compare the age of the universe predicted by this
model with the constraints on the age coming from the oldest
globular cluster observations. We find that the best estimated
values for the age of the universe is $14.95 \pm 0.42$ Gyr that is
in agreement with the constraint in the age coming from the oldest
globular clusters.

The present model has been constrained using SNe Ia data. It means
that we have constrained the model using information (observations)
of the past of the universe until a redshift $z\lesssim 1.5$. So,
SNe Ia observations are not enough to trace the whole evolution of
the universe from the Big-Bang until the present time. Therefore, it
would be important to constrain the model using also cosmological
observations that can provide information of the early universe like
the observations of the CMB anisotropies  from the WMAP experiment
and the large scale structure (LSS) from SDSS experiment. A
preliminary work in that sense has been already done in
\cite{ArturoUlisesProc1} showing (but not in a conclusive way) that
when the model is constrained using the shift parameter $R$ of CMB
and the parameter $A$ measuring the BAO peak from SDSS, in addition
to the Gold 2006 SNe Ia data set, the estimated value of
$\tilde{\zeta}$ is negative at 99.7\% confidence level, violating
the second law of thermodynamics and the estimated value of $H_0$ is
lower ($H_0 \sim 53 ({{\rm km}}/{{\rm s}}){{\rm Mpc}}^{-1}$) than
that reported by 5 year WMAP and HST Cepheid variable star
observations and giving a bad goodness-of-fit to data. Considering
these facts we may conjecture that this model does not work well for
early times of the universe and that in order to be a viable model
the bulk viscosity should be triggered just until late times.

Another problem of the model is to explain the origin of the  bulk
viscosity from known or new physics. Some proposals in that sense
have been explored, for instance in \cite{Wilson} and
\cite{Mathews2008}, where it is proposed a mechanism to generate the
bulk viscosity by the decay of dark matter particles into
relativistic products.

Finally, we explore also the sensitivity of the results to different
ways of marginalization over $H_0$ when we estimate just the
coefficient $\tilde{\zeta}$. We find that  it is almost negligible
the difference in the estimations of $\tilde{\zeta}$ when $H_0$ is
marginalized assuming \emph{constant} and \emph{Gaussian} priors
centered in the values reported by 5 year WMAP and HST Cepheid
variable star observations. However, when $H_0$ is marginalized by
assuming a specific value (a Dirac delta prior located at the values
reported as mentioned above) the estimated values of $\tilde{\zeta}$
have a small increase and the values of $\chi^2_{\rm{d.o.f.}}$
increase as well (i.e., we have a worse fit to data).

        \section{Acknowledgements.}

We thank Jose Antonio Gonz{\'a}lez for his comments on the
constraints for the age of the universe, Francisco S. Guzm{\'a}n and
Olivier Sarbach for useful discussions during the preparation of
this work. We also thank John Barrow for his useful comments on
previous works done on this topic. We acknowledge to the Instituto
Avanzado de Cosmolog\'{\i}a (IAC) for its partial support and useful
seminars. This work was in part supported by grants SNI-20733,
CIC-UMSNH No. 4.8,  UMSNH-CA-22 and COECyT--FIFOECYT 2008.


                \appendix

        \section{Marginalization assuming a constant prior over $H_0$.}
        \label{SectionAnalyticalMargMethod}

To construct a \textbf{pdf} that depends only on the parameter $\tilde{\zeta}$ we use the process
of \emph{marginalization} over $H_0$ in order to eliminate the dependence of the \textbf{pdf} with
respect to the parameter $H_0$.

The following procedure can be applied for any other model. It is, in a model with several
parameters where it is necessary to reduce the number of free parameters or, for some reason, to
eliminate the dependance of the model from some particular free parameters (for instance: $H_0$) in
order to compute the best estimated values of the other free parameters that we keep (in our case:
$\tilde{\zeta}$) the solution is to \emph{marginalize} over the parameters that we want to eliminate.


In the present work we assume three different prior distribution
functions for $H_0$: constant, Gaussian and Dirac delta. In this
appendix we describe the marginalization using a constant prior for
$H_0$.

We start by multiplying and dividing by $H_0$ the equation \eref{luminosity_distance1}
\begin{equation}\label{luminosity_distanceAppendixFDM}
d_L(z, \tilde{\zeta}, H_0) = \frac{c(1+z)}{H_0} \int_0^z \frac{H_0 \; dz'}{H(z',
\tilde{\zeta}, H_0)}
\end{equation}
Next, we define a new dimensionless luminosity distance as $D_L(z, \tilde{\zeta}) \equiv H_0
\cdot
d_L(z, \tilde{\zeta}, H_0) / c$. Then
\begin{equation}
    D_L(z, \tilde{\zeta}) = c(1+z) \int_0^z \frac{dz'}{E(z', \tilde{\zeta})}
\end{equation}
\noindent where $D_L(z, \tilde{\zeta})$ \textit{does not depend} on $H_0$ anymore and
$E(z, \tilde{\zeta}) \equiv H(z, \tilde{\zeta}, H_0)/H_0$.

The theoretical distance moduli \eref{distanceModuli} becomes
\begin{equation}\label{distanceModuli2}
\fl \mu^{\rm{t}}(z, \tilde{\zeta}, H_0)=5\log_{10} \left( \frac{D_L(z, \tilde{\zeta}) \cdot c}
{H_0 \cdot \rm{Mpc}} \right) + 25 \equiv 5\log_{10} \left(
\frac{D_L(z, \tilde{\zeta})}{\tilde{H}_0}\right) + 25
\end{equation}

\noindent where we have defined a dimensionless ``Hubble parameter'' $\tilde{H}_0 \equiv
H_0 \cdot {\rm Mpc} \, / c$. It is useful to define a new theoretical distance moduli that
does not depend on $H_0$ anymore as
\begin{equation}\label{DistanceModuliwithoutHo}
\tilde{\mu}^{\rm{t}}(z, \tilde{\zeta}) \equiv 5 \log_{10}[D_L(z, \tilde{\zeta})] + 25
\end{equation}

\noindent Then, the expression \eref{distanceModuli2} for the distance moduli becomes
\begin{equation}\label{equationA5appendix}
\mu^{\rm{t}}(z, \tilde{\zeta}, H_0)=\tilde{\mu}^{\rm{t}}(z, \tilde{\zeta}) - 5
\log_{10}[\tilde{H}_0]
\end{equation}
Now we construct the $\chi^2$ function \eref{ChiSquareDefinition} with these new definitions as

\begin{equation}\label{chi2NewAppendix}
\chi^2(\tilde{\zeta}, H_0)= \sum_{i=1}^n \left( \frac{\tilde{\mu}^{\rm{t}}(z_i, \tilde{\zeta})
- \mu^{\rm{obs}}_i - 5 \log_{10} \tilde{H}_0}{\sigma_i} \right)^2
\end{equation}

\noindent where $\mu^{\rm{obs}}_i$ is the observed distance moduli and $\sigma_i$ its variance.

We rewrite the expression \eref{chi2NewAppendix} as
\begin{eqnarray}
 \chi^2(\tilde{\zeta}, H_0) =& \sum_{i=1}^n \left( \frac{\tilde{\mu}^{{\rm
t}}_i-\mu_i^{{\rm obs}}}{\sigma_i} \right)^2 - 2 \left(5
\log_{10}\tilde{H}_0 \right)  \sum_{i=1}^n \left(
\frac{\tilde{\mu}^{{\rm t}}_i-\mu_i^{{\rm obs}}}{\sigma^2_i} \right)   \nonumber \\
& + \left(5\log_{10}\tilde{H}_0 \right)^2 \sum_{i=1}^n \left(
\frac{1}{\sigma^2_i} \right)
\label{chiHugeAppendix}
\end{eqnarray}
If we define \footnote{Note that these expressions do not depend on
$H_0$ anymore.}
\begin{equation}\label{ABC}
A \equiv \sum_{i = 1}^n \left( \frac{\tilde{\mu}^{{\rm t}}_i -
\mu^{{\rm obs}}_i}{\sigma_i} \right)^2, \qquad B \equiv \sum_{i =
1}^n  \frac{\tilde{\mu}^{{\rm t}}_i - \mu^{{\rm
obs}}_i}{\sigma_i^2}, \qquad C \equiv \sum_{i= 1}^n
\frac{1}{\sigma_i^2}
\end{equation}

\noindent then we can express \eref{chiHugeAppendix} as
\begin{equation}\label{Chi2ABCx}
 \chi^2(\tilde{\zeta}, H_0) = A-2Bx+Cx^2
\end{equation}
\noindent where
\begin{equation}\label{xequaltoHo}
x\equiv 5 \log_{10} (\tilde{H}_0).
\end{equation}
Note that all the dependence of the $\chi^2$ function with respect to $H_0$ is now in the $x$
variable. Remind that the \textbf{pdf} is defined as
\begin{equation}\label{expChi}
{{\rm {\bf pdf}}}(\tilde{\zeta}, \tilde{H}_0) = cte \cdot {{\rm e}}^{-\chi^2 \,/2}
\end{equation}
\noindent where ``{\it cte}'' is a normalization constant. So, we marginalize the \textbf{pdf}
\eref{expChi} over $\tilde{H}_0$ computing  the following integration

\begin{equation}\label{MarginalizationIntegral2}
 {{\rm {\bf pdf}}}(\tilde{\zeta}) = \int^{\infty}_{-\infty}
 {{\rm {\bf pdf}}}(\tilde{\zeta}, \tilde{H}_0) \cdot {{\rm {\bf pdf}}}
 (\tilde{H}_0) \; d\tilde{H}_0
\end{equation}

\noindent where ${{\rm {\bf pdf}}} (\tilde{H}_0)$ is the
\textit{prior} probability density function for $\tilde{H}_0$. In
general, the integration \eref{MarginalizationIntegral2} has to be
done over the range of all the possible values of the parameter to
be marginalized. In this case the range of possible values for
$H_0$, or equivalently $\tilde{H}_0$, is $(-\infty, \infty)$.

If we take the case where the  prior for $\tilde{H}_0$ is a \emph{constant} then:

\begin{equation}\label{MarginalizationIntegral1}
 {{\rm {\bf pdf}}}(\tilde{\zeta}) = cte \int^{\infty}_{-\infty}
 {{\rm {\bf pdf}}}(\tilde{\zeta}, \tilde{H}_0) \; d\tilde{H}_0
\end{equation}

A \emph{constant} prior means that we do not prefer any particular
value for $\tilde{H}_0$, i.e., any value for $\tilde{H}_0$ has the
same probability of being. To solve the integral
\eref{MarginalizationIntegral1} we do a change of variable from
$\tilde{H}_0$ to $x$ (see expression \eref{xequaltoHo}). At the same
time, we substitute the expression \eref{Chi2ABCx} into
\eref{expChi} and then \eref{expChi} into
\eref{MarginalizationIntegral1}, yielding

\begin{eqnarray}\label{pdf1}
\fl {{\rm {\bf pdf}}}(\tilde{\zeta}) = cte \; \left( \frac{\ln10}{5}\right) \,{{\rm
exp}}\left[\frac{1}{2} \left(\frac{\tilde{B}^2}{C} -A\right) \right]
\int^{\infty}_{-\infty} {{\rm exp}} \left[-\frac{C}{2} \left( x -
\frac{\tilde{B}}{C} \right)^2 \right]dx
\end{eqnarray}

\noindent where $\tilde{B} \equiv B + (\ln 10) / 5 $.

We can see that the integral \eref{pdf1} has the form of a Gaussian distribution with value:
\begin{equation}
 1 = \frac{1}{ \sigma \sqrt{2 \pi}} \int^{\infty}_{-\infty} {{\rm exp}}
 \left[ - \frac{(x-\bar{x})^2}{2 \sigma^2} \right] dx
\end{equation}

\noindent where $\bar{x}$ is the mean and $\sigma^2$ is the variance of the distribution.

Therefore the expression \eref{pdf1} becomes
\begin{equation}
{{\rm {\bf pdf}}}(\tilde{\zeta}) = cte \; \left( \frac{\ln10}{5}\right) \sqrt{\frac{2\pi}{C}} \,
{{\rm exp}}\left[-\frac{1}{2} \left(A-\frac{\tilde{B}^2}{C} \right)\right]
\end{equation}

\noindent so that the ${\rm {\bf pdf}}(\tilde{\zeta})$ does not depend on $H_0$ anymore. Note that
it was not necessary any \emph{numerical} integration of the expression
\eref{MarginalizationIntegral1}.
We can express this ${\rm {\bf pdf}}(\tilde{\zeta})$ in terms of a new $\chi^2_{\rm cp}$ function
like

\begin{equation}
{\rm {\bf pdf}}(\tilde{\zeta})= a \cdot {\rm e}^{-\chi^2_{\rm cp}/2}
\end{equation}

\noindent where $a \equiv cte \cdot \sqrt{2\pi} \ln10 / (5\sqrt{C})$ and

\begin{equation}\label{chi_AnalyticalIntegration}
\chi^2_{{\rm cp}}(\tilde{\zeta}) \equiv A(\tilde{\zeta}) -  \frac{\left[B(\tilde{\zeta}) +
\ln(10)/5 \right]^2}{C}
\end{equation}

\noindent This new $\chi^2_{\rm cp}$ function does not depend on $H_0$ anymore. The label ``cp''
stands for \textit{constant prior} for $H_0$.

        \section{Marginalization assuming a Gaussian prior over $H_0$.}
        \label{SectionNumericalGaussMargMethod}

We perform the marginalization assuming a \emph{Gaussian} probability distribution function for
$H_0$ centered at $H^{\ast}_0$ and with standard deviation $\sigma^{\ast}$. So, the \textbf{pdf}
($H_0$) prior with the form of a Gaussian distribution for $H_0$ is
\begin{equation}
{{\rm {\bf pdf}}}(H_0)=\exp\left[ -\frac{1}{2} \left(\frac{H_0 - H^{\ast}_0}{\sigma^{\ast}}
\right)^2 \right]
\end{equation}

\noindent With this, the expression \eref{MarginalizationIntegral2} becomes
\begin{equation}\label{MargGaussNumericalHoDefinition}
{\rm {\bf pdf}}(\tilde{\zeta}) = cte \cdot \int_{-\infty}^\infty
\rme^{-\chi^2 / 2}\; \exp\left[ -\frac{1}{2} \left(
\frac{H_0 - H^{\ast}_0}{\sigma^{\ast}} \right)^2 \right] d H_0
\end{equation}

\noindent where the $\chi^2 $ function is given by \eref{Chi2ABCx} and ``{\it cte}'' is a
normalization constant.

In the present work we use two different central values for the Gaussian prior coming from two
different observations. One of them, is that coming from the 5 year WMAP observations
where $H^{\ast}_0=70.5 \,({{\rm km}}/{{\rm s}}) {{\rm Mpc}}^{-1}$ with an standard deviation of
$\sigma^{\ast}=1.3 \,({{\rm km}}/{{\rm s}}) {{\rm Mpc}}^{-1}$. The other central value is
$H^{\ast}_0=72 \,({{\rm km}}/{{\rm s}}) {{\rm Mpc}}^{-1}$ with an standard deviation of
$\sigma^{\ast}=8 \,({{\rm km}}/{{\rm s}}) {{\rm Mpc}}^{-1}$ as reported by the HST Cepheid variable
star observations.

In practice we perform the numerical integration of the expression
\eref{MargGaussNumericalHoDefinition} in the interval
 $H_0=[55,85]\,({{\rm km}}/{{\rm s}}) {{\rm Mpc}}^{-1}$ considering that this is a suitable and
representative interval that include almost the 100\% of the probability density for $H_0$.

        \section{Marginalization assuming a Dirac delta prior over $H_0$.}
        \label{SectionDiracDeltaPriorMargMethod}

It means to assume a specific value of $H_0$. It is like to think that the value of the Hubble
constant is $H_0^{\ast}$ that has been measured with
an infinite accuracy (with standard deviation equal to zero), so that its \textbf{pdf}($H_0$) has
the form of a \emph{Dirac delta} function.

Clearly, this assumption is just an idealization to simplify the work and it does not correspond to
the reality. However, for some cases it is a good approximation.
This prior has the advantage of that once assumed it is very simple to perform the integration of
the expression \eref{MarginalizationIntegral2} using the Dirac delta properties.
Thus, the prior with the form of a  Dirac delta for $H_0$ is
\begin{equation}
{{\rm {\bf pdf}}}(H_0)=\delta(H_0-H^{\ast}_0)
\end{equation}

\noindent With this, the expression \eref{MarginalizationIntegral2} becomes
\begin{equation}\label{PDFMargDiracPrior}
{\rm {\bf pdf}}(\tilde{\zeta}) = cte \cdot \rme^{-\chi^2(\tilde{\zeta}, H_0^{\ast}) / 2}
\end{equation}

\noindent where ``{\it cte}'' is a normalization constant.

\end{document}